%% file: dlaff.tex
\documentclass{article}
\usepackage{amsmath,amsfonts,amsthm}
\usepackage{graphicx,url,calc}

\usepackage[lined,ruled,vlined]{algorithm2e}

\usepackage{multirow}
\usepackage{rotating}
\usepackage{xspace}

\newtheorem{thm}{Theorem}[section]

\newtheorem{lem}[thm]{Lemma}
\newtheorem{cor}[thm]{Corollary}

\newtheorem{rem}[thm]{Remark}

\newcommand{\GO}{{\cal O}}
\newcommand{\pF}[1]{\leavevmode
        \kern.1em\raise.0ex \hbox{\Z}\kern-.1em /\kern-.15em\lower.3ex
         \hbox{#1}\lower.3ex \hbox{\Z}}

\newcommand{\Zp}{\leavevmode\kern.1em\raise.0ex \hbox{\ensuremath{\mathbb{Z}}}\kern-.1em /\kern-.15em\lower.3ex\hbox{p}\mbox{\ensuremath{\mathbb{Z}}}\xspace}
\newcommand{\Z}{\ensuremath{\mathbb{Z}}\xspace}

\newcommand{\linbox}{{\sc LinBox}}
\newcommand{\MM}{\ensuremath{\text{MM}}\xspace}

\newcommand{\LQUP}{\ensuremath{\text{LQUP}}\xspace}
\newcommand{\TRSM}{\ensuremath{\text{TRSM}}\xspace}
\newcommand{\LTL}{\ensuremath{\text{LTL}}\xspace}
\newcommand{\UTUT}{\ensuremath{\text{UTUT}}\xspace}
\newcommand{\UTLT}{\ensuremath{\text{UTLT}}\xspace}
\newcommand{\INVT}{\ensuremath{\text{INVT}}\xspace}

\newcommand{\ADD}{\ensuremath{\text{ADD}}\xspace}
\newcommand{\lup}{\ensuremath{\text{LUP}}\xspace}

\newcommand{\lqup}{\ensuremath{\text{LQUP}}\xspace}
\newcommand{\turbo}{\ensuremath{\text{TURBO}}\xspace}

\newcommand{\tu}{\turbo}
\newcommand{\trsm}{\texttt{trsm}\xspace}
\newcommand{\dtrsm}{\texttt{dtrsm}\xspace}
\newcommand{\ftrsm}{\texttt{ftrsm}\xspace}
\newcommand{\dbl}{\texttt{double} }

\newcommand{\til}{\lower 2pt\hbox{\small${}^\sim$}}

\newcommand{\lCeil}{\left\lceil}
\newcommand{\rCeil}{\right\rceil}
\usepackage{fancyhdr}
\title{Dense Linear Algebra over Word-Size Prime Fields: the FFLAS and
  FFPACK packages\footnote{This material is based on work supported in
    part by the Institut de Math\'ematiques Appliqu\'ees de Grenoble,
    project IMAG-AHA. This work was mostly done while the second author was a postdoctoral fellow of the Symbolic Computation Group, D.R. Cheriton School 
of Computer Science, University of Waterloo, Canada.}}
            
\urldef\jgdemail\url{Jean-Guillaume.Dumas@imag.fr}
\urldef\pgemail\url{Pascal.Giorgi@lirmm.fr}
\urldef\cpemail\url{Clement.Pernet@imag.fr}

\author{Jean-Guillaume Dumas\footnote{Laboratoire Jean Kuntzmann, umr
    CNRS 5224, 51, rue des Math\'ematiques BP 53 IMAG-LMC,
F38041 Grenoble, France; \jgdemail}\\ Universit\'e de Grenoble 
  \and Pascal Giorgi\footnote{Laboratoire d'Informatique de Robotique
    et de Microélectronique de Montpellier, umr CNRS 5506; \pgemail}\\ Universit\'e de Montpellier  
  \and Cl\'ement Pernet\footnote{MOAIS (INRIA Rh\^one-Alpes / CNRS LIG
    Laboratoire d'Informatique de Grenoble); \cpemail}\\  Universit\'e de Grenoble}
           
\begin{document}
\maketitle       

\begin{center}{\bf Abstract}\\[10pt]
\begin{minipage}{\textwidth*5/6}\small
In the past two decades, some major efforts have been made to reduce
exact (e.g. integer, rational, polynomial) linear algebra problems 
to matrix multiplication in order to provide algorithms with optimal asymptotic complexity. 
To provide efficient implementations of such algorithms one need to be careful with the underlying arithmetic.
It is well known that  modular techniques such as the Chinese remainder algorithm or the $p$-adic lifting allow 
very good practical performance, especially when word size arithmetic are used.
Therefore, finite field arithmetic becomes an important core for efficient exact linear algebra libraries.
In this paper, we study high performance implementations of basic linear algebra routines
over word size prime fields: specially the matrix multiplication; our goal being to provide an exact alternate to the numerical BLAS library.
We show that this is made possible by a careful combination of numerical computations and asymptotically faster algorithms.
Our kernel has several symbolic linear algebra applications enabled by diverse 
matrix multiplication reductions: symbolic triangularization,
 system solving, determinant and matrix inverse implementations are thus studied.
\end{minipage}\end{center}
      

\newpage
{\small
\tableofcontents
}
\input{intro}


\input{preliminaries}

\input{fflas}

%
\input{trsm}
%

\input{ffpack}

%
\input{concl}

\appendix

\input{winobound}

\bibliographystyle{plain}
\addcontentsline{toc}{section}{References}
\bibliography{jgdbibl}

\end{document}

%% file: intro.tex
\section{Introduction}\label{sec:intro}
Finite fields play a crucial role in computational algebra. Indeed,
finite fields are the basic representation used to solve many integer
problems. The whole solutions are then gathered via the Chinese
remainders or lifted p-adically. Among those problems are integer polynomial
factorization \cite{Zassenhaus:1978:RHF}, integer system solving
\cite{Dixon:1982:Pad,Storjohan:2005:HighOrder}, integer matrix normal 
forms
\cite{jgd:2001:jsc} or integer determinant \cite{Kaltofen:2005:CCDet}.
Finite fields are of intrinsic use in polynomial linear 
algebra \cite{Giorgi:2003:issac} but also 
in cryptology (e.g. large integer 
factorization \cite{Montgomery:1995:BLA}, discrete logarithm 
computations \cite{Odlyzko:2000:DLPF}) or for error correcting codes.
Moreover, nearly all of these problems involve linear algebra 
resolutions.
Therefore, a fundamental issue is to implement efficient elementary 
arithmetic
operations and very fast linear algebra routines over finite fields.

We propose a way to implement the equivalent of
the basic BLAS level 1, 2, and 3 numerical routines 
(respectively dot product,
matrix-vector product and matrix-matrix product), but 
over finite fields.
We will focus on implementations over fields with small
cardinality, namely not exceeding machine word size,
but with any characteristic (consequently, we do not deal 
with
optimizations for powers of 2 cardinalities). For instance, 
we show that {\it symbolic matrix multiplication can be as fast
  as numerical matrix multiplication} (see section \ref{sec:mm}) when using word
size prime fields.
Our aim is {\em not} to rebuild some specialized routines for each
field instance. Instead, the main
idea is to use a very efficient and automatically tuned
numerical library as a kernel (e.g. ATLAS \cite{Whaley:2001:AEO})
and to make some conversions in order to perform 
an {\em exact} matrix multiplication
(i.e. {\em without any loss of precision}). The efficiency will be
reached by performing as few conversions as possible. 
Several alternatives to this approach exist: one would be to implement
a core linear algebra with integer arithmetic. Unfortunately, new
architectures focus on numerical arithmetic and therefore by using integer
arithmetic we would lose a factor of 2 or 4 due to the SIMD (single
instruction, multiple data) SSE speed-up of the numerical
routines. Note that SSE4 with some integer support is announced for 2008
  and might then change some of this point of view. Anyway, another
  feature of our approach is to rely on a large community of effort
  for the numerical handling of linear algebra routines. We want to show in this
  paper that no real gain could be obtained by trying to mimic their effort
  over just using it.

Then, building on this fast numerical blocks, we
can use fast
matrix multiplication algorithms, such as Strassen's or Winograd's
variant \cite[\S 12]{VonzurGathen:1999:MCA}.
There, we use exact computation on a higher level
and therefore do not suffer from instability
problems \cite{Higham:1990:EFM}. 

Many algorithms have been designed to use matrix multiplication in order to be able to prove an optimal theoretical complexity.
In practice those exact algorithms are only seldom used.
This is the case, for example, in many linear algebra problems such as determinant, rank, inverse, system solution or minimal and characteristic polynomial. 
We believe that with our kernel, each one of those optimal complexity algorithms can also be the most efficient. 
One goal of this paper is then to show the actual effectiveness of this belief. 
In particular we focus on factorization of matrices of any shape and any rank. 

Some of the ideas from preliminary versions of this paper \cite{jgd:2002:fflas}, in particular the BLAS-based matrix multiplication for small prime fields, are now incorporated 
into the Maple computer algebra system since its version 8 and also into the 2005 version of the computer algebra system Magma. 
Therefore an effort towards effective reduction has been made \cite{jgd:2004:ffpack} in C++ and within Maple by A. Storjohann\cite{Storjohann:2003:ACA}. 
Effective reduction for minimal and characteristic polynomial were proposed in \cite{jgd:2005:charp} and A. Steel has reported on similar efforts 
within his implementation of some Magma routines. 

In this paper, the matrix factorization, namely the exact equivalent of the LU factorization is thus extensively studied. 
Indeed, unlike numerical matrices, exact matrices are very often singular, even more so if the matrix is not square~!
Consequently, Ibarra, Moran and Hui have developed generalizations of the LU factorization, namely the LSP and LQUP factorizations
\cite{Ibarra:1982:LSP}. Then we adapt this scheme to rank, determinant, inverse (classical or Moore-Penrose), nullspace computations, etc. 
There, we will give not only the asymptotic complexity measures but the constant factor of the dominant term. 
Most of these terms will give some constant factor to the multiplication time and we will compare those theoretical ratios to
the efficiency that we achieve in practice. This will enable us to give a measure of the effectiveness of our reductions (see especially
section \ref{sec:use}).

Now, we provide a full C++ package available
directly \cite{DumGioPer:2006:ffpack} 
or through the exact linear algebra library
 \linbox\footnote{\scriptsize\texttt{www.linalg.org}} \cite{jgd:2002:icms}.
Extending the work undertaken 
by the authors et
al.\cite{Pernet:2001:Winograd,jgd:2002:fflas,Brassel:2003:eccad,Giorgi:2003:ACA,jgd:2004:dotprod,jgd:2004:ffpack,jgd:2005:charp},
this
paper focuses on matrix multiplication with an extended Winograd variant
optimizing memory allocation ; on simultaneous triangular system solving; on matrix factorization and improved constant
factors of complexity for many linear algebra equivalent routines (inverse, squaring, upper-lower or upper-upper triangular multiplication, etc.).

The paper is organized as follows. 
Section \ref{sec:prem} introduces some material for the evaluation of arithmetical costs of recursive algorithms;
we also motivate our choice to represent elements of a finite field; 
Then section \ref{sec:mm} presents efficient ways to implement  matrix multiplication over {\em generic} prime fields, 
including a study of fast matrix multiplication. Section \ref{sec:trsm} deals
with the matrix multiplication based simultaneous resolution of $n$
triangular systems. 
Laslty, section \ref{sec:pack} presents implementations of several matrix factorizations
and their applications with a study of complexity and of efficiency in
practice.
%

%% file: preliminaries.tex
\section{Preliminaries}\label{sec:prem}

\subsection{Finite field arithmetic}

The first task, to implement exact linear algebra routines, 
is to develop the underlying arithmetic.
Indeed, any finite field, except $GF(2)$, do not map
directly to the arithmetical units of nowadays processors and  a
software emulation is therefore mandatory. 
This has been well studied in literature, and we refer to
\cite{jgd:2004:dotprod} and references therein for a survey on this topic. 
Here, we recall the different ways of implementing 
such arithmetic and we will motivate our choice of a particular one
for efficient linear algebra routines.

\subsubsection{Implementations}

Representation of finite fields elements plays a crucial role in the
efficiency of arithmetic operations. From now on, we will count
arithmetic operations in  terms of field operations, that is
we will count addition, subtraction, multiplication and division in
the arithmetic complexity results.

A usual way to implement prime fields arithmetic is to map the elements of the field to integers modulo a prime number, 
defined by its characteristic. From now on, we will focus on prime fields with characteristic no greater than a word size (e.g. 32 bits).
In this basic case, various representations and arithmetics can be used:
\begin{itemize}
\item {\bf Classical representation with integer divisions}.\\
Integers between $0$ and $p-1$ or between $(1-p)/2$ and $(p-1)/2$ are
used; additive group operations are 
done with machine integers operations followed by a test and a correction; multiplication is followed by machine remaindering
while division is performed via the extended gcd algorithm.
\item {\bf Montgomery representation}.\\
This representation, proposed in \cite{Montgomery:1985:MMT}, allows to avoid costly machine remaindering within the multiplication.
A shifted representation is used and remaindering is replaced by multiplications.
Note that others operations, except the division, stay identical.
\item {\bf Floating point inverse}.\\
Another idea to reduce remaindering cost in multiplication is to precompute the inverse of the characteristic $p$ within
a floating point number. Therefore, only two floating point multiplications and some rounding
are necessary. However, floating point rounding may induce a $\pm 1$ error and then an adjustment is required, as implemented
in Shoup's NTL library \cite{Shoup:NTL}.
\item {\bf Discrete logarithm (also called Zech logarithm)}.\\
Here, elements are seen as a power of a generator of the multiplicative group, namely a primitive element.
As a consequence, multiplicative group operations can be performed only by addition or subtraction modulo $p-1$.
Nevertheless, this representation makes the addition/subtraction more complicated in the field.
In particular, these operations need some table lookup; see \cite[\S2.4]{jgd:2004:dotprod}.
\end{itemize}

Extension fields, denoted $GF(p^k)$, are usually implemented via polynomials over the prime field \Zp
modulo an irreducible polynomial of degree $k$. Thus, operations in
the extension reduce to polynomial arithmetic. An alternative is to
tabulate entries and use the Zech logarithm representation also.
As for prime fields, some representations can be used to avoid the costly remaindering phase within the multiplication.
We will not discuss any implementations over extension field in this
paper. We let the reader refer to 
\cite{jgd:2007:dqt} for details on data structures, arithmetic 
and matrix multiplication over small extension fields.
From now on, when we will refer to finite fields this will mean
word-size prime
fields and the extensions for which the trick of 
\cite[\S 4]{jgd:2002:fflas} is usable.



\subsubsection{Ring homomoprphism and delayed reduction}\label{ssec:ffperf}

As a primitive tool for implementing linear algebra routines, the
efficiency of the finite field representation needs to be well studied.
In \cite{jgd:2004:dotprod} the author analyzes the efficiency of finite field arithmetic according to a chosen representation.
It has been shown that atomic operations (e.g. addition,
multiplication) can be performed more efficiently than with the classic method 
depending on the architecture.
In particular, it appears that memory access based implementations (i.e. discrete logarithm) and
floating point based implementations (i.e. floating point inverse) are
more efficient on older architecture such as Ultra Sparc.
Nevertheless, with newer architecture such as Pentium III and Pentium 4, integer machine operations become more efficient and outperform
other implementations, except discrete logarithm  for multiplicative group operations.

However, for linear algebra, the primary operation is the succession of two operations: a multiplication followed by an addition; this operation 
is commonly called {\it AXPY} (also ``fused-mac'' or FMA within hardware). 
This operation clearly influences the efficiency of vectors dot product which is one of the main operations of classic linear algebra.
However, 
optimized {\it AXPY} atomic operation
is deprecated since one would rather use delayed divisions.
This technique consists in successive multiplications and accumulations without any division.
Divisions intervene either just before an overflow occurs within the hardware data, or 
 only after a fixed numbers of accumulations.

Indeed, any prime field $\Z_p$ can be naturally embedded into $\Z$ by representing its elements
 with an integer of an interval $[m, M]$, such that $M-m = p-1$.
 The reverse conversion consists in applying a reduction modulo $p$ to the integer
 value.

 The ring structure being preserved by these homomorphisms, any ring algorithm over
 $\Z_p$ can be transposed into a ring algorithm over $\Z$.

 Now the machine integer arithmetic uses a fixed number of bits $\gamma$ for the integer
 representation: $\gamma = 32$ for \texttt{int}, $\gamma=24$ (resp. $\gamma=53$) for
 single (resp. double) precision floating point values, etc.

 Using this approximate integer arithmetic, one has therefore to ensure that the
 computation of the integer algorithm 
 will not overflow the representation. Hence for each integer algorithm, a bound on the maximal
 computed value has to be given, depending on $m$ and $M$.

For example, if the representation is interval is $[0,p-1]$, 
 one can perform $\lambda$ accumulations without any divisions if
\begin{equation}\label{eq:fond}
\lambda(p-1)^2 < 2^\gamma \leq (\lambda+1)(p-1)^2
\end{equation}
Note that if signed words are available, a centered
representation can be used (i.e. $-\frac{p-1}{2} \leq x \leq
\frac{p-1}{2}$ for the storage of an element $x$ of the odd prime field) 
and the equation \ref{eq:fond} becomes
\begin{equation}\label{eq:centered}
\lambda \left( \frac{p-1}{2}\right)^2 < 2^{\gamma-1} \leq (1+\lambda)\left( \frac{p-1}{2}\right)^2
\end{equation}
which improves $\lambda$ by a factor of $2$.

Hence, the bottleneck of divisions can be amortized since only 
$\left\lceil \frac{n}{\lambda}\right\rceil$ divisions will occur 
in a $n$-dimensional vector dotproduct.

Contrary to atomic operations, floating point based implementations
for dotproduct 
tend to be the most efficient on average.
In particular, timings are constant and achieve almost half of the
peak of arithmetical unit while the timings of others implementations drop 
as soon as the size of the finite field increases.
However, when small primes are used, one can improve these timings
to almost the peak of the machine by using others implementations \cite[\S3.4]{jgd:2004:dotprod}.

According to these results and the necessity of genericity,
we provide implementations based on generic finite fields (e.g. use of
{\it C++ template mechanism}). 
However, in this paper, we mainly use a floating point based implementation for our finite fields arithmetic, called {\tt Zpz-double}.
This choice is principally  motivated by the use of optimized
numerical basic linear algebra operations through the BLAS library.
Indeed, one can easily benefit from these libraries by simply mapping linear algebra operations over finite fields
to numeric computations and delayed divisions. This will be extensively explained in sections \ref{sec:mm} and \ref{sec:trsm}.
Therefore, the choice of floating point based representations for finite field elements will be an asset since it will 
avoid any data conversion. Possibly, we may use a different finite
field implementation in order to compare efficiencies. 
There, we will use the notation {\tt Zpz-int}, meaning a word size integer based implementation.
As we will see throughout the rest of the paper, the combination of BLAS and {\tt Zpz-double} implementation 
 will allow us to approach numerical efficiency for linear algebra problems over finite fields.

\subsection{Recursion materials for arithmetical complexity} 

The following two lemmas will be useful to study the constant factor of linear algebra 
algorithms compared to matrix multiplication.
The first one gives the order of magnitude when the involved matrices
will be square:
\begin{lem}\label{lem:m}
  Let $m$ be a positive integer and suppose
  that \begin{enumerate}
  \item $T(m) = C T(\frac{m}{2}) +
    am^\omega  + \epsilon(m)$, with $\epsilon(m) \leq  g m^2 $ for some
    constants $C,a,\omega,g$.
  \item $T(1) = e$ for some constant $e$.
  \item $\log_2( C)  < \omega$.
  \end{enumerate}
  Then $T(m) = \GO(m^\omega)$.
\end{lem}
\begin{proof}
  Let $t=\log_2(m)$.
  The recursion gives,
  $$T(m)=C^t T(1) + am^\omega \frac{1-
    \left(\frac{C}{2^\omega}\right)^t}{1-\frac{C}{2^\omega}}
  + \sum_{i=0}^{t-1} C^i \epsilon( \frac{m}{2^i} ). 
  $$
  Then, on the one hand, if $C \neq 4$ this yields 
  $T(m) = \frac{a2^\omega}{2^\omega-C} m^\omega + k C^t + g' m^2$,
  where $g' <\frac{4g}{4-C}$ and 
  $k < T(1)-\frac{a2^\omega}{2^\omega-C}-g'$.
  On the other hand, when $C=4$, we have 
  $T(m)= \frac{a2^\omega}{2^\omega-C} m^\omega + k' C^t + gm^2
  \log_2(m)$,
  where $k'< T(1)-\frac{a2^\omega}{2^\omega-C}$.
  In both cases, with $C^t=m^{\log_2(C)}$, this gives 
  $T(m)=\frac{a2^\omega}{2^\omega-C} m^\omega +o(m^\omega)$.
\end{proof}
Now we give the order of magnitude when the matrix dimensions differ:
\begin{lem}\label{lem:mn}
  Let $m$ and $n$ be two positive integers and suppose
  that \begin{enumerate}
  \item $T(m,n) = \sum_{i=1}^k c_i T(\frac{m}{2},n-d_i\frac{m}{2}) +
    am^\omega + b m^{\omega-1}n + \epsilon(m,n)$, with $C=\sum_{i=1}^k
    c_i$, $D=\sum_{i=1}^k c_i d_i$, $2<\omega$ and $ \epsilon(m,n) \leq  g m^2 + h mn$ .
  \item $T(1,F) \leq e F$ for a constant $e$.
  \item $\log_2( C)  < \omega - 1$
  \end{enumerate}
  Then $T(m,n) = \GO(m^\omega+m^{\omega-1}n)$.
\end{lem}
\begin{proof}
  As in the preceding lemma, we use the recursion and geometric sums to
  get
  \begin{multline}
  T(m,n) = \sum_{i_1=1}^k c_{i_1} \ldots  \sum_{i_t=1}^k c_{i_t}
  T(1,n-f(d_1,\ldots,d_t,m)) + \\
  m^\omega \left(a \frac{1-
      \left(\frac{C}{2^\omega}\right)^t}{1-\frac{C}{2^\omega}}
    -bD \frac{1-
      \left(\frac{C}{2^{\omega-1}}\right)^t}{1-\frac{C}{2^{\omega-1}}}
  \right)
  + bm^{\omega-1}n  \frac{1-
    \left(\frac{C}{2^{\omega-1}}\right)^t}{1-\frac{C}{2^{\omega-1}}}\\
 + \sum_{i_1=1}^k c_{i_1} H(m/2,n-d_i m/2) \ldots 
 +  \sum_{i_1=1}^k c_{i_1} \ldots  \sum_{i_t=1}^k c_{i_t}
  H(1,n-f(d_1,\ldots,d_t,m))
   \end{multline}
Thus, we get $$\alpha m^\omega+\beta m^{\omega-1}n \leq T(m,n) \leq
\alpha  m^\omega+\beta m^{\omega-1}n + C^t T(1,n) + \sum_{i=1}^t C^i
H(\frac{m}{2^i},n).$$ The last term is bounded by 
$gm^2 \frac{1-
    \left(\frac{C}{4}\right)^t}{1-\frac{C}{4}} + fmn\frac{1-
    \left(\frac{C}{2}\right)^t}{1-\frac{C}{2}}$ when 
$C\neq 4$ and $C \neq 2$. In this case
$C^t T(1,n) + \sum_{i=1}^t C^i H(\frac{m}{2^i},n) \leq 
m^{\log_2(C)}\left( (e+\frac{2g}{C-2}) n+ \frac{4g}{C-4}\right) = \GO(
m^\omega+m^{\omega-1}n)$. When $C=2$, a supplementary $\log_2(m)$
factor 
arises in the small factors, but the order of magnitude is preserved
since $\log_2( C)  +1 = 2 < \omega$.
\end{proof}
These two lemmas are useful in the following sections where we 
solve (e.g.suppose $T(m)=\alpha m^\omega$ in a recurring relation for
$\alpha$)
to get the actual constant of the
dominant term. Thus, when we give an equality on
complexities, this equality means that the dominant terms of both
complexities are equal.
In particular, some lower order terms may differ.

%% file: fflas.tex
\section{Matrix multiplication}\label{sec:mm}








We propose a design for a matrix multiplication kernel routine over a word-size
finite field, based on the three following features:
\begin{enumerate}
  \item delayed modular redution, as explained section \ref{ssec:ffperf},
  \item cache tuning and floating point arithmetic optimizations using BLAS,
  \item Strassen-Winograd fast algorithm.
\end{enumerate}

\subsection{Cache tuning using BLAS}
In most of the modern computer architecture, a memory access to 
the RAM is more than one hundred times slower than an 
arithmetic operation. To circumvent this slowdown, the memory 
is structured into two or three levels of cache acting as buffers 
to reduce the number of accesses to the RAM and reuse as much 
as possible the buffered data. 
This approach is only valid if the algorithm involves many computations with
local data. 

In linear algebra, matrix multiplication is the better suited operation for
 cache optimization: it is the first basic operation, 
for which the time complexity $\GO(n^3)$ is an order of magnitude 
higher than the space complexity $\GO(n^2)$. Furthermore it plays such a central
role in linear algebra, that every other algorithm will take advantage of the
tuning of this kernel routine.

These considerations have driven the development of basic 
linear algebra subroutines (BLAS) \cite{Dongarra:1990:BLAS,Whaley:2001:AEO} for numeric
computations. One of its main achievement is the level 3 set of routines,
based on a highly tuned matrix multiplication kernel. 

For computations on a word-size finite field, a similar approach could be
developed, e.g. following \cite{Gustavson:1998:RBD} for block decomposition. 
Instead, we propose to simply wrap these numerical routines to form
the integer algorithm of the delayed modular approach of the previous
section. This will enable to take benefit from both the efficiency of the
floating point arithmetic and the cache tuning of the BLAS
libraries. Furthermore relying on the generic BLAS interface makes it possible
to benefit from the large variety of optimizations for all existing architectures
and ensures a long term efficiency thanks to the much larger development effort
existing for numerical computations.
 
\begin{figure}[htbp]\begin{center}
\includegraphics[width=0.95\textwidth,height=180pt]{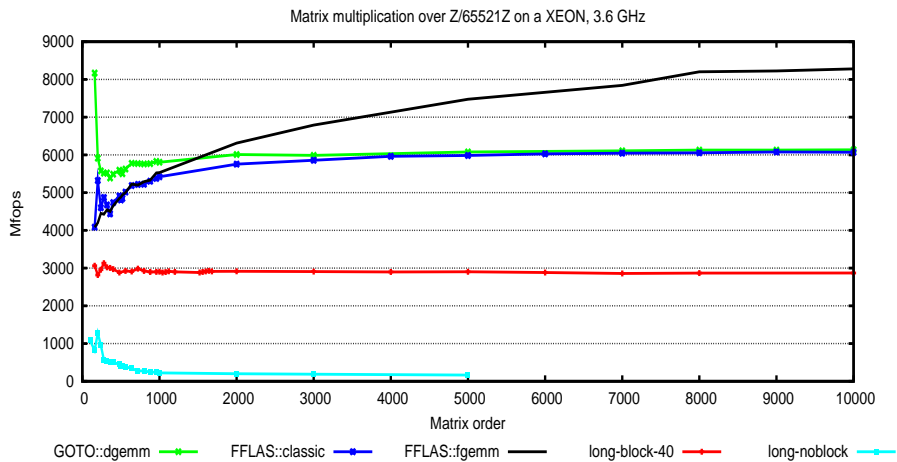}
\caption{Blocking classical matrix multiplication, on a Xeon, 3.6GHz.}
\label{fig:bloc}
\end{center}\end{figure}
Figure \ref{fig:bloc} shows the advantage of this method
(\texttt{FFLAS::classic}) compared to two other
implementations: the naive algorithm (\texttt{long-noblock}), and a hand-made cache
tuned implementation, based on block decomposition of the input matrices, so
that each block product  could be performed locally in the L2 cache memory
(\texttt{long-block-40}, for a block dimension $40$). 
The graph compares the computation speed in millions of field operations per
seconds (Mfops) for different matrix orders.
As a comparison we also provide the computation speed
of the equivalent numerical BLAS routine \texttt{dgemm}. 
This approach improves on the efficiency of the two other methods over a finite
field and the overhead of the modular reductions is limited.
Finally, the (\texttt{FFLAS::fgemm}) implementation is the most efficient thanks to the combination of numerical computations
 and a fast matrix multiplication algorithm which is discussed in the next section.

\subsection{Winograd fast algorithm}\label{ssec:winograd}
The third feature of this kernel is the use of a fast matrix
multiplication algorithm. 
We will focus on Winograd's variant \cite[algorithm 12.1]{VonzurGathen:1999:MCA} of Strassen's
algorithm \cite{Strassen:1969:GENO}.
We denote by $\MM(n)$ the dominant term of the 
arithmetic complexity of the
matrix multiplication. The value of $\MM(n)$ thus reflects the choice
of algorithm, e.g. $\MM(n)=2n^3$ for the classical
algorithm, and mean that the actual complexity of the classical
algorithm is $2n^3+\GO(n^2)$. 
We also denote by $\omega$ the asymptotic exponent of $\MM(n)$,
it is thus $3$ for the classical algorithm, $\log_2(7)\approx
2.807354922$ for the Strassen-Winograd
variant, and the best known exponent is about $2.375477$ by
\cite{Coppersmith:1990:MMAP}. 

In \cite{Higham:1990:EFM} Winograd's variant is discarded for
numerical computations because of its bad stability and despite its
better running time. In \cite{Kaporin:2004:TCS}
aggregation-cancellation techniques of \cite{Laderman:1992:PAA} are also
compared. They also give better stability than the Winograd variant
but  worse running time. For exact computation, stability is no longer an issue
and Winograd's faster variant is thus preferred.

\subsubsection{A Cascade structure}\label{ssec:cascade}
Asymptotically, this algorithm improves on the number of arithmetic
operations required for matrix multiplication from
$\MM(n)=2n^3$ 
to $\MM(n)=6n^{2.8074}$. 
But for a given $n$, the total number of arithmetic operations can be reduced by
switching after a few recursive levels of Winograd's algorithm to the classic
algorithm.
Table \ref{tab:winolevel} compares the number of arithmetic operations depending
on the matrix order and the number of recursive levels.
\begin{table}[htbp]
\begin{center}
\begin{tabular}{|c|c|c|c|c|c|c|c|c|}
\cline{3-8}
\multicolumn{2}{c}{}&\multicolumn{6}{|c|}{Recursive levels of Winograd's algorithm}\\
\hline
$n$&Classic& 1 &2 & 3 & 4 & 5 & 6 \\
\hline
4   & \textbf{112}& 144     & 214     &         &         &         & \\
8   & \textbf{960}     & 1024    & 1248    & 1738    &         &         &\\
16  & 7936    &\textbf{7680}& 8128    & 9696    & 13126   &         & \\
32  & 64512   & 59392   & \textbf{57600}   & 60736   & 71712   & 95722   &\\
64  & 520192  & 466944  & 431104  & \textbf{418560}  & 440512  & 517344  & 685414\\
\hline
\end{tabular}
\caption{Number of arithmetic operations in the multiplication of two $n\times n$ matrices} \label{tab:winolevel}
\end{center}
\end{table}

This phenomenon is amplified by the fact that additions in 
classic matrix multiplication are cheaper than the ones in Winograd algorithm
since they take advantage of the cache optimization of the BLAS routine. 
As a consequence, the optimal number of recursive levels depends on the
architecture and must be determined experimentally. It can be described by a simple
parameter: the matrix order $w$ for which one recursive level is as fast 
the classic algorithm. Then the number of levels $l$ is given by the formula~·
$$
l = \left\lfloor \text{log}_2 \frac{n}{w}\right\rfloor+1.
$$
\subsubsection{Schedule of the algorithm}

We based our implementation of Winograd's algorithm on two different schedules.
For the operation  $C \leftarrow A\times B$ we use that of \cite[Fig. 1]{Douglas:1994:gemmw} and 
for the extended $C \leftarrow \alpha A\times B +\beta C$, that of
\cite[Fig. 6]{Huss-Lederman:1996:mai} that we recall in table
\ref{tab:schedule:ABC}.
More details about tasks scheduling and memory efficient variants of Winograd's algorithm can be
found in \cite{DumasPernet:2007:WinoSchedule}.
\begin{table}[htbp]
\begin{center}
\begin{tabular}{|llr|llr|}
\hline
\# & operation & loc. & \#° & operation & loc.  \\  
\hline
1 & $S_1 = A_{21} + A_{22}$	& $X_1$	& 12 & $S_4 = A_{12} - S_2$	& $X_1$ \\ 
2 & $T_1 = B_{12} - B_{11}$	& $X_2$	& 13 & $T_4 = T_2 - B_{21}$	& $X_2$ \\
3 & $P_5 = \alpha S_1 T_1 $	& $X_3$	& 14 & $C_{12} =  \alpha S_4 B_{22} + C_{12}$ & $C_{12}$ \\
4 & $ C_{22} = P_5 + \beta C_{22}$	& $C_{22}$ & 15 & $U_5 = U_2 + C_{12} $	& $C_{12}$\\
5 & $ C_{12} = P_5 + \beta C_{12}$	& $C_{12}$ & 16 & $P_4 = \alpha A_{12}T_4-\beta C_{21}$ & $C_{21}$\\ 
6 & $S_2 = S_1 - A_{11}$	& $X_1$	& 17 & $S_3 = A_{11} - A_{21}$	& $X_1$   \\  
7 & $T_2= B_{22} - T_1$	& $X_2$	& 18 & $T_3 = B_{22} - B_{12}$	& $X_2$   \\
8 & $P_1 = \alpha A_{11} B_{11}$	& $X_3$	& 19 & $U_3 = \alpha S_3 T_3+U_2$& $X_3$\\  
9 & $C_{11} = P_1 + \beta C_{11}$	& $C_{11}$ & 20 & $U_7 = U_3 + C_{22}$	& $C_{22}$ \\
10& $U_2 = \alpha S_2 T_2 + P_1$	& $X_3$	& 21 & $U_6 = U_3 - C_{21}$	& $C_{21}$ \\
11&$U_1 =\alpha A_{12}B_{21}+C_{11}$	& $C_{11}$ &  & &\\
 \hline
\end{tabular}
\caption{Schedule for operation $C\leftarrow \alpha A\times B+\beta C$ with 3 temporaries}
\label{tab:schedule:ABC}
\end{center}
\end{table}

\subsubsection{Control of the overflow}

Since Winograd's algorithms will be used with delayed modular reductions,
one has to ensure that any intermediate computation will fit in the underlying fixed-size
integer representation being used. 
Indeed, intermediate values can become large in this algorithm, and the former
bound for the dot-product no-longer holds. 

The main result of this section is that, in the worst case, the largest
intermediate computation occurs during the recursive 
computation of the sixth recursive product $P_6$ (see appendix
\ref{app:winobound}). 
This result generalizes \cite[theorem 3.1]{jgd:2002:fflas}
for the computation of $AB+\beta C$.

\begin{thm}\label{th:bound}
Let $A \in \Z^{m \times k}$, $B \in \Z^{k \times n}$  $C \in \Z^{m
  \times n}$ be three matrices and $\beta \in \Z$ with
$ m_A \leq a_{i,j} \leq M_A $, $m_B \leq b_{i,j} \leq M_B$ and  $m_C \leq
c_{i,j} \leq M_C$. Moreover, suppose that $0\leq -m_A \leq M_A$,  $0\leq -m_B \leq
M_B$, $0\leq -m_C \leq M_C$, $M_C \leq M_B$ and $|\beta| \leq M_A, M_B$.
Then every intermediate value $z$ involved in the computation of $A \times B +
\beta C$ with $l$ ($l\geq 1$) recursive levels of Winograd algorithm satisfy:

\[ \left|z \right| \leq
  \left(\frac{1+3^l}{2}M_A+\frac{1-3^l}{2}m_A\right)\left(\frac{1+3^l}{2}M_B+\frac{1-3^l}{2}m_B\right)
  \left\lfloor{ \frac{k}{2^l}}\right\rfloor  
\]
Moreover, this bound is optimal.
\end{thm}

The proof is given in appendix \ref{app:winobound}.

Using a positive integer representation of the prime field elements (integers between $0$
and $p-1$), the following corollary holds:

\begin{cor}[Positive modular representation] \label{cor:positif}
Using the same notations, with 
$a_{i,j},b_{i,j},c_{i,j},\beta \in [0\dots p-1]$, we have
\[ \left|z \right| \leq \left(\frac{1+3^l}{2}\right)^2 \left\lfloor{
    \frac{k}{2^l}}\right\rfloor (p-1)^2 \]
\end{cor}


Instead, using a balanced representation (integers between $-\frac{p-1}{2}$ and
$\frac{p-1}{2}$), this bound can be improved:

\begin{cor}[Balanced modular representation]\label{cor:centre}
Using the same notations with
$a_{i,j},b_{i,j},c_{i,j},\beta \in [-\frac{p-1}{2}\dots\frac{p-1}{2}]$,
we have
\[ \left|z \right| \leq \left(\frac{3^l}{2}\right)^2 \left\lfloor{
    \frac{k}{2^l}}\right\rfloor \left(p-1\right)^2 \]
\end{cor}


\begin{cor}\label{cor:winokmax}
One can compute $l$ recursive levels of Winograd algorithm without
modular reduction over integers of $\gamma$ bits as long as 
$k<k_\text{Winograd}$ where
$$
k_\text{Winograd} =
\left(\frac{2^{\gamma+2}}{\left((1+3^l)(p-1)\right)^2}+1\right)2^l
$$
for a positive modular representation and
$$
k_\text{Winograd} =
\left(\frac{2^{\gamma+2}}{\left(3^l(p-1)\right)^2}+1\right)2^l
$$
for a balanced modular representation.

\end{cor}
%
%
%
\subsection{Timings and comparison with numerical routines} \label{ssec:fgemm-perf} 

This section presents experiments of our implementation of the matrix
multiplication kernel described above.

The experiments use two different BLAS library: the automatically tuned BLAS
ATLAS \cite{Whaley:2001:AEO}, and the BLAS by Kazushige Goto
\cite{2002:gotoblas} refered to as GOTO. We used the \verb!gcc!
compiler version 4.1 on the Xeon machine and the \verb!icc! compiler
version 9.0 on the Itanium.
We recall that \texttt{dgemm} refers to the BLAS matrix multiplication routine
over double precision floating point numbers. Similarly, we named 
our routine over a word-size finite field \texttt{fgemm}.

%
%

\begin{table}[htb]\begin{center}
\begin{tabular}{|cc|c||r|r|r|r|r|r|r|r|r|}
\cline{3-11}
\multicolumn{2}{c|}{} & $n$          
& {\em 1000}  & {\em 2000}  & {\em 3000}  & {\em 5000} & {\em 7000}  & {\em 8000} & {\em 9000}& {\em 10000}\\
\cline{3-11}
\multicolumn{11}{c}{}\\[-0.2cm]
\hline
& & fgemm  
& $0.38$s    &  $2.73$s    & $8.59$s    &
$36.34$s &   $95.21$   & $134.03$s  & $190.21$s & $258.08$ \\ 
\cline{3-11}
& & dgemm 
&  $0.37$s     &  $2.98$s    & $10.02$s   &
$46.10$s &  $126.38$s   & $188.97$s  & $267.83$s & $368.30$s \\ 
\cline{3-11} \\[-.3cm]
\cline{3-11}
& \begin{rotate}{90}\scriptsize ATLAS \end{rotate} & {\bf
  $\frac{fgemm}{dgemm}$} 
& \bf 1.02  & \bf 0.92 & \bf 0.86
&  \bf 0.79 &  \bf 0.75 & \bf 0.71 & \bf 0.71 & \bf 0.70 \\   
\hline
\multicolumn{11}{c}{}\\[-0.2cm]
\hline

& & fgemm 
& $0.36$s    &  $2.53$s    & $7.95$s    &
$33.44$s &  $87.46$s   & $124.86$s  & $177.25$s & $238.00$s \\ 
\cline{3-11}
& & dgemm 
& $0.34$s    &  $2.65$s    & $8.90$s   &
$41.01$s &   $112.31$s   & $167.20$s  & $237.16$s & $325.62$s\\ 
\cline{3-11} \\[-.3cm]
\cline{3-11} & \begin{rotate}{90}\scriptsize GOTO \end{rotate} & {\bf
  $\frac{fgemm}{dgemm}$}  
& {\bf 1.05} & \bf 0.96 & \bf
0.89  &  \bf 0.82 &  \bf 0.78 & \bf 0.75 & \bf 0.75 & \bf 0.73 \\   

\hline
\multicolumn{11}{c}{}\\[-0.1cm]
\end{tabular}
%
\caption{Comparison between \texttt{fgemm} and \texttt{dgemm} on a  Xeon, 3.6GHz}\label{tab:matmulp4}
\end{center}
\end{table}
%
%
%

%
%
%
%
%
\begin{table}[htb]\begin{center}
\begin{tabular}{|cc|c||r|r|r|r|r|r|r|r|}
\cline{3-11}
\multicolumn{2}{c|}{} & $n$          & {\em 1000}  & {\em 2000}  & {\em 3000}  & {\em 5000} & {\em 7000}  & {\em 8000} & {\em 9000} & {\em 10000}\\
\cline{3-11}
\multicolumn{11}{c}{}\\[-0.2cm]
\hline
& & fgemm  & $0.46$s & $3.22$s & $10.14$s & $42.28$s & $110.64$s & $163.53$s & $225.08$s & $296.56$s  \\
\cline{3-11}
& & dgemm & $0.45$s & $3.49$s & $11.45$s & $53.12$s & $144.45$s & $215.53$s & $305.21$s & $419.00$s  \\
\cline{3-11} \\[-.3cm]
\cline{3-11}
& \begin{rotate}{90}\scriptsize ATLAS \end{rotate} & {\bf
  $\frac{fgemm}{dgemm}$}  & \bf 1.01 & \bf 0.92 & \bf 0.89 & \bf 0.80 & \bf 0.77 & \bf 0.76 & \bf 0.74 & \bf 0.71 \\
\hline
\multicolumn{11}{c}{}\\[-0.2cm]
\hline
& & fgemm  & $0.43$s & $2.99$s & $9.35$s & $39.21$s & $104.07$s & $152.12$s & $209.22$s & $277.32$s  \\
\cline{3-11}
& & dgemm  & $0.40$s & $3.18$s & $10.61$s & $48.88$s & $133.75$s & $200.11$s & $283.94$s & $390.37$s  \\
\cline{3-11} \\[-.3cm]
\cline{3-11}
 & \begin{rotate}{90}\scriptsize GOTO \end{rotate} & {\bf $\frac{fgemm}{dgemm}$}  & \bf 1.06 & \bf 0.94 & \bf 0.88 & \bf 0.80 & \bf 0.78 & \bf 0.76 & \bf 0.74 & \bf 0.71 \\
\hline
\end{tabular}
\caption{Comparison between \texttt{fgemm} and \texttt{dgemm} on Itanium2, 1.3GHz}\label{tab:matmulia64}
\end{center}
\end{table}

The tables \ref{tab:matmulp4} and \ref{tab:matmulia64} report timings obtained
for both exact and numeric matrix multiplication. First the comparison shows
that the exact computation over a word size finite field (modulo 65521 on these tables) 
can reach a similar
range of efficiency as the numerical computation.
For increasing matrix dimensions, the exact computation becomes even more
efficient (see also figure \ref{fig:bloc}), thanks to the use of Winograd's algorithm (improvement factor between
$13\%$ and $29\%$ for dimension $10\,000$).



These experiments also show the advantage of relying on a generic interface for
numerical BLAS: the exact computation will directly take advantage of the
improvements of the best numerical routine. This appears when comparing GOTO and
ATLAS on these two target architecture, where GOTO is about $10\%$ faster.




%% file: trsm.tex
\section{Triangular system solving with matrix right/left hand side}\label{sec:trsm}
We now discuss the implementation of solvers for 
triangular systems with matrix right hand side (or equivalently left 
hand side). 
%
The resolution of such systems plays a central role in many linear algebra
problems, e.g. it is the second main operation 
in block Gaussian elimination after matrix multiplication as will be recalled in section \ref{sec:triang}. This operation is commonly named 
\trsm in the BLAS convention. In the following, we will consider
without loss of generality the resolution of an upper triangular system with
matrix right hand side, i.e. the operation $B \leftarrow U^{-1}B$, where $U$ is
$m\times m$ upper triangular and $B$ is $m\times n$.

Following the approach of the BLAS numerical routine,
our implementation is based on  a block recursive algorithm 
to reduce the computation to matrix multiplications. 

Now similarly to our approach with matrix multiplication, the design of our
implementation also focuses on delaying the modular reductions as
much as possible. As will be shown in section \ref{ssec:trsmdel}, delaying the
whole resolution leads to a quick growth in the size of coefficients.
Therefore we also present in section \ref{ssec:trsmdelupdate} another way of
delaying these modular reductions.
We lastly present how to combine these two techniques within a multi-cascade
algorithm.



\subsection{The block recursive algorithm}\label{ssec:rec-trsm}


Algorithm \texttt{trsm} recalls the block recursive algorithm.

\begin{algorithm}
\dontprintsemicolon
\caption{\trsm($A,B$)}\label{alg:trsm:rec}
\KwData{ $A \in \Zp^{m \times m}$, $B \in \Zp^{m \times n}$.}
\KwResult{$X \in \Zp^{m \times n}$ such that $AX=B$.}
\Begin{
\eIf{$m=1$}{
 $ X:= A_{1,1}^{-1} \times B$\;
}{
 \tcc{splitting matrices into two blocks of sizes $\left\lfloor \frac{m}{2}
  \right\rfloor$ and $\lCeil \frac{m}{2} \rCeil$ 
\[ 
\begin{array}{cccc}
A & X & & B \\
\overbrace {\left[ \begin{array}{cc} A_1 & A_2 \\ & A_3 \end{array} \right] }&
\overbrace{\left[ \begin{array}{ccc} & X_1 & \\ & X_2 & \end{array} \right] }&
= &
\overbrace{\left[ \begin{array}{ccc} & B_1 & \\ & B_2 & \end{array} \right] }
\end{array}
\]}

$X_2:=$\trsm($A_3,B_2$)\;
$B_1:= B_1 - A_2X_2$\;
$X_1:=$\trsm($A_1,B_1$)\;
}
}
\end{algorithm}

\begin{lem}\label{lem:trsm}
Algorithm \trsm\ is correct and the leading term of its arithmetic
complexity over $\Zp$ is 
$$\TRSM(m,n) = 
    \frac{1}{2^{\omega-1}-2}\lCeil\frac{n}{m}\rCeil  \MM(m) 
$$
This complexity is 
$m^2n$
using classic matrix  multiplication.
\end{lem}

\begin{proof}
Extending the previous notation \MM(n), we denote by \MM(m,k,n) the cost of
multiplying a $m\times k$ by a $k\times n$ matrices.
The cost function $\TRSM(m,n)$ satisfies the following equation:
$$\TRSM(m,n)= 2\TRSM(\frac{m}{2},n)+\MM(\frac{m}{2},\frac{m}{2},n).$$
Let $t=\log_2(m)$. Although the algorithm works for any $n$, we restrict the
complexity analysis to the case where $m \leq n$ for the sake of simplicity.
We then have:
\begin{eqnarray*}
\TRSM(m,n)&=& 2\TRSM(\frac{m}{2},n)+
\frac{1}{2^{\omega-1}}\lCeil\frac{n}{m}\rCeil \MM(m)
\\&=& 2^t \TRSM(1,n) + \frac{1}{2^{\omega-1}}\lCeil\frac{n}{m}\rCeil \MM(m)
\frac{ 1 -
  \left(\frac{2}{2^{\omega-1}}\right)^t}{1-\frac{2}{2^{\omega-1}}}.
\end{eqnarray*}
As $\TRSM(1,n)=2n$ and $\left(2^{\omega-1}\right)^t = m^{\omega-1}$,
we obtain the expected complexity 
$\TRSM(m,n)=\frac{1}{2^{\omega-1}-2}\lCeil\frac{n}{m}\rCeil \MM(m) + \GO(m^2+mn).$
\end{proof}

\subsection{Delaying reductions globally}\label{ssec:trsmdel}

As for matrix multiplication, the delayed computation 
relies on the fact that ring operations over the
finite field can be replaced by ring operations over \Z using the ring
homomorphisms described in section \ref{ssec:ffperf}.
However, triangular system resolutions involve, in the general case, field
operations: the divisions by the diagonal elements of the triangular matrix.
Therefore this technique is only valid with unit diagonal matrices.

In the general case, the triangular matrix is made unit diagonal by the
following factorization: $A=DU$, where $D$ is diagonal and $U$ is unit diagonal
upper triangular. Then the system $U X = D^{-1}B$ only involves ring operations
and can be solved over \Z.
This normalization leads to an additional cost of $O(mn)$ arithmetic
operations (see \cite{jgd:2004:ffpack} for more details).

Now the integer computation with a fixed sized arithmetic (e.g. the floating point
arithmetic)  is exact as long as all intermediate results of the computation
do not exceed the bit capacity of the representation.
Therefore we now propose bounds on the values computed by the algorithm over \Z.


%
\begin{thm}  \label{th:trsmbound}
Let $T \in \mathbb{Z}^{ n\times n}$ be a unit diagonal upper triangular matrix and $b \in \mathbb{Z}^n$, 
with $m \leq T_{i,j} \leq M$ and $m \leq b_i \leq M$ and  $m\leq 0\leq M$.
Let $x = ( x_i )_{i \in [1 \dots n]} \in \mathbb{Z}^n$ be the solution of the
system $Tx=b$.
Then $\forall \ k \in [0\dots n-1]$~:
\[ \left \{
\begin{array}{ll}
 -u_k \leq x_{n-k} \leq v_k &  \text{for $k$ even,}\\ 
 -v_k \leq x_{n-k} \leq u_k &  \text{for $k$ odd}
\end{array}
\right. 
\]
with 
\[
\left \{
\begin{array}{l}
u_k =\frac{M-m}{2}(M+1)^k  - \frac{M+m}{2}(M-1)^k,\\
v_k = \frac{M-m}{2}(M+1)^k  + \frac{M+m}{2}(M-1)^k. \\
\end{array}
\right. 
\]
\end{thm}
\begin{proof}

First note the following relations:
$$
\forall k \left\{
\begin{array}{lcl}
u_k &\leq& v_k \\
-mu_k &\leq &Mv_k\\
-mv_k &\leq &Mu_k\\
\end{array}
\right.
$$
The third one comes from
$$
Mu_k+mv_k=\frac{M^2-m^2}{2}((M+1)^k-(M-1)^k) \geq 0.
$$
The proof is now an induction on $k$, following the system resolution order.
The initial case $k=0$ correspond to the first step:
$x_n=b_n$, leading to
$$ -u_0 = m \leq x_n \leq M = v_0.$$ 
Suppose now that the inequalities hold for $k\in [0\dots l]$
and prove them for  $k=l+1$.
If  $l$ is odd, $l+1$ is even.
{\small
\begin{eqnarray*}
x_{n-l-1}&=& b_{n-l-1} - \sum_{j=n-l}^n{T_{n-l-1,j}x_j}\\
&\leq& M + \sum_{i=0}^{\frac{l-1}{2}}{\max(Mu_{2i},-mv_{2i}) + \max(Mv_{2i+1},-mu_{2i+1})}\\
&\leq& M\left(1 + \sum_{i=0}^{\frac{l-1}{2}}{u_{2i} + v_{2i+1}}\right)  \\ 
&\leq& M\left(1 + \sum_{i=0}^{\frac{l-1}{2}}{\frac{M-m}{2}(M+2)(M+1)^{2i} +
\frac{M+m}{2}(M-2)(M-1)^{2i}} \right)  \\ 
&\leq& M\left(1 + \frac{M-m}{2}(M+2)\frac{(M+1)^{l+1}-1}{(M+1)^2-1} +
\frac{M+m}{2}(M-2)\frac{(M-1)^{l+1}-1}{(M-1)^2-1} \right)\\  
&\leq&\frac{M-m}{2}(M+1)^{l+1}  + \frac{M+m}{2}(M-1)^{l+1} = v_{l+1}.
\end{eqnarray*}
Similarly,
\begin{eqnarray*}
x_{n-l-1}  &\geq& m - \sum_{i=0}^{\frac{l-1}{2}}{\max(Mv_{2i},-mu_{2i}) +
\max(Mu_{2i+1},-mv_{2i+1})}\\ 
&\geq& m -  M\sum_{i=0}^{\frac{l-1}{2}}{v_{2i} + u_{2i+1}}\\ 
&\geq& m -M\sum_{i=0}^{\frac{l-1}{2}}{\frac{M-m}{2}(M+2)(M+1)^{2i} -
\frac{M+m}{2}(M-2)(M-1)^{2i} }  \\ 
&\geq& m - M\left(\frac{M-m}{2}(M+2)\frac{(M+1)^{l+1}-1}{(M+1)^2-1} -
\frac{M+m}{2}(M-2)\frac{(M-1)^{l+1}-1}{(M-1)^2-1} \right)\\  
&\geq&\frac{M-m}{2}(M+1)^{l+1}  - \frac{M+m}{2}(M-1)^{l+1} = u_{l+1}.
\end{eqnarray*}
}
For $l$ even, a similar proof leads to
$$
-v_{l+1} \leq x_{n-l-1} \leq u_{l+1}.
$$
\end{proof}
\begin{cor}\label{cor:trsmoptimal}
 Using the notation of theorem \ref{th:trsmbound}, 
$$ 
 |x| \leq \frac{M-m}{2}(M+1)^{n-1}  + \frac{M+m}{2}(M-1)^{n-1}.
$$
Moreover this bound is optimal.
\end{cor}
\begin{proof} 
The sequence $(v_k)$ is increasing and always greater than $(u_k)$. 
Thus $\forall \ k \in [0\dots {n-1}] \ |x_{n-k}|\leq \ u_k \leq v_k \leq v_{n-1}$.

Now the vector $x = ( x_i )_{i \in [1\dots n]} \in \mathbb{Z}^n$ such that 
$ \forall \ k \in [0\dots n-1] \ |x_{n-k}| = v_k$ satisfies the system $Tx=b$ with
$$
T = \
\left[
\begin{array}{ccccc}
\ddots & \ddots & \ddots & \ddots & \ddots \\
       &   1    &  M   &    m   & M  \\
       &        &    1   &   M  & m   \\
       &        &        &    1   & M \\
       &        &        &        & 1   \\
\end{array}
\right], 
b = \left[\begin{array}{c}\vdots\\m\\M\\m\\M \end{array} \right]
$$
Therefore the bound is reached.
\end{proof}
The following corollaries apply this result to the positive and balanced modular
representations.

\begin{cor}[Positive modular representation]\label{cor:trsmpositif}
For $1 \leq i,j \leq n$, if  $T_{i,j},b_i \in [0\dots p-1]$, then 
$$
|x| \leq \frac{p-1}{2}(p^{n-1}  + (p-1)^{n-1}).
$$ 
\end{cor}
\begin{cor}[Balanced modular representation]\label{cor:trsmcentre}
For $1 \leq i,j \leq n$, if  $T_{i,j},b_i \in [-\frac{p-1}{2}\dots
\frac{p-1}{2}]$, then
$$
|x| \leq \frac{p-1}{2}\left(\frac{p+1}{2}\right)^{n-1}.
$$ 
\end{cor}

\begin{rem}
The balanced modular representation improves the bound by a factor of $2^{n-1}$.
\end{rem}

As a consequence, one can solve a unit diagonal triangular system of dimension
$n$ using arithmetic operations with integers stored on $\gamma$ bits if
\begin{equation}\label{eq:trsmboundpos}
\frac{p-1}{2}(p^{n-1}  + (p-1)^{n-1})< 2^{\gamma}
\end{equation}
for a positive representation and 
\begin{equation}\label{eq:trsmboundcen}
\frac{p-1}{2}\left(\frac{p+1}{2}\right)^n< 2^{\gamma}
\end{equation}
for a balanced representation.

For instance, using the  \dbl floating point representation ($53$ bits of
mantissa)
the maximal dimension of the system is $34$ (resp. $52$) for a positive
(resp. balanced) representation of $\Z_3$. 
For larger fields, this maximal dimension becomes quickly very small: with
$p=1001$, $n\leq 5$ (resp. $n\leq 6$) for a positive (resp. balanced)
representation. 

In the following, we will denote by $t_\text{del}(p,\gamma)$ the maximum
dimension for the resolution with delayed modular reductions.
This dimension is small, and this approach can therefore only be used 
as a terminal case of the recursive block algorithm. 
This first cascade algorithm is characterized by the threshold 
$t_\text{del}$.
For efficiency, we used in our implementation the BLAS routine \trsm to perform
the delayed computation over \Z.
Despite the small dimension of the blocks, we will see in section
\ref{ssec:trsmexp} that this approach can slightly improve the efficiency of the
computation when the finite field is small. 

\subsection{Delaying reductions in the update phase only} \label{ssec:trsmdelupdate}
The block recursive algorithm consists in several matrix multiplications of
different dimensions. In most cases, the matrix multiplications are done over \Z
with a modular reduction on the result only. But part of these result matrices
will be accumulated to other matrix multiplications in later computations.
Therefore these intermediate modular reductions  could be delayed even more by 
allowing to accumulate these results over \Z as much as possible.

This technique can be applied within the former cascade algorithm, to produce a
double cascade structure. The key idea is to split the matrices at two levels as
shown on figure \ref{fig:trsm:recblasdelayed}: 
\begin{figure}[htbp]\begin{center}
\includegraphics[width=0.8\textwidth]{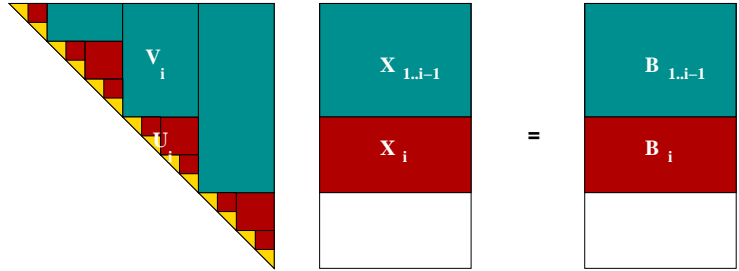}
\caption{Splitting for the double cascade \trsm algorithm}
\label{fig:trsm:recblasdelayed}
\end{center}\end{figure}
a fine grain
splitting with the dimension $t_\text{del}$ of the previous section, and a
coarse grain splitting with the dimension $t_\text{update}$ such that
all recursive calls of dimension lower than $t_\text{update}$ can let the 
 matrix multiplication updates accumulate without modular reductions. 
Choosing $t_\text{update} = k_\text{Winograd}$ (from corrolary \ref{cor:winokmax}) 
will ensure this property.
To adjust together the dimensions of the two block decompositions, we set
$t_\text{split} = \left \lfloor t_\text{Winograd} / t_\text{del} \right
\rfloor t_\text{del}$.
\begin{algorithm}
\dontprintsemicolon
\caption{\texttt{trsm-rec-BLAS-delayed}~:} 
\label{alg:trsm:recblasdelayed}
\KwData{ $A \in \Zp^{m \times m}$, $B \in \Zp^{m \times n}$}
\KwResult{$X \in \Zp^{m \times n}$ s.t. $AX=B$}
\Begin{
Compute $t_\text{del}$ from equation (\ref{eq:trsmboundpos} or \ref{eq:trsmboundcen}) \;
Compute $t_\text{Winograd}$ from corrolary (\ref{cor:winokmax}) \;
$t_\text{split} = \left \lfloor t_\text{Winograd} / t_\text{del} \right
\rfloor t_\text{del}$\;
\ForEach{block column of $A$ of dimension $m\times t_\text{split}$ of the form
$\begin{bmatrix}V_i\\U_i\\0\end{bmatrix}$}{
$X_i = \texttt{trsm-partial-delayed} (U_i,B_i)$ \;
$X_i = X_i \mod p$\;
$B_{1\dots i-1} = B_{1\dots i-1} - V_i X_i$\;
$B_{1\dots i-1} = B_{1\dots i-1} \mod p$\;
}
\Return{$X$}
}
\end{algorithm}
\begin{algorithm}
\dontprintsemicolon
\caption{\texttt{trsm-partial-delayed}}
\label{alg:trsmdelaye}
\KwData{ $A \in \Zp^{m \times m}$, $B \in \Zp^{m \times n}$, $m$ must be lower
  than  $t_\text{update}$}
\KwResult{$X \in \Zp^{m \times n}$ s.t. $AX=B$}
\Begin{
\eIf{ $m\leq n_\text{del}$}{
$B=B \mod p$\;
$X = \texttt{dtrsm}(A,B)$ \tcc*{the BLAS routine}\;
$X=X \mod p$\;
}{
\tcc{ (splitting of the matrix into blocks of dimension $\left\lfloor \frac{m}{2}
 \right\rfloor$ and $\lCeil \frac{m}{2} \rCeil$) }\;
$
\begin{array}{cccc}
A & X & & B \\
\overbrace {\left[ \begin{array}{cc} A_1 & A_2 \\ & A_3 \end{array} \right] }&
\overbrace{\left[ \begin{array}{ccc} & X_1 & \\ & X_2 & \end{array} \right] }&
= &
\overbrace{\left[ \begin{array}{ccc} & B_1 & \\ & B_2 & \end{array} \right] }
\end{array}
$\;
$X_2:=\texttt{trsm-partial-delayed} (A_3,B_2)$ \;
$B_1:= B_1 - A_2X_2$  \tcc*{without modular reduction} \;
$X_1:=\texttt{trsm-partial-delayed} (A_1,B_1)$ \;
}
\Return $X$
}
\end{algorithm}

Algorithm \ref{alg:trsm:recblasdelayed} is a loop on every block of column dimension
$t_\text{update}$. For each of them, the triangular system is solved using algorithm
\ref{alg:trsmdelaye} and the update is performed  by a matrix multiplication over
\Z followed by a modular reduction.
Algorithm \ref{alg:trsmdelaye} is simply the cascade algorithm of the previous
section: the block recursive algorithm \ref{alg:trsm:rec} with the fully delayed
algorithm as a terminal case.
The matrix multiplication updates are performed over \Z without any reduction of
the result, since the threshold $t_\text{update}$ allows to accumulate them.


\subsection{Experiments}
\label{ssec:trsmexp}

We now compare three implementations of the \trsm routine over a word size finite
field: 
\begin{description}
\item Pure recursive (\texttt{Pure-Rec}): Simply algorithm \ref{alg:trsm:rec},
\item Recursive-BLAS  (\texttt{Rec-BLAS}): The cascade algorithm formed by
  the recursive algorithm and the BLAS routine \dtrsm as a terminal case. It
  differs from algorithm \ref{alg:trsmdelaye} by the fact that the matrix
  multiplication $B_1:= B_1 - A_2X_2$  is always followed by a modular reduction.
\item Recursive-BLAS-Delayed (\texttt{Rec-BLAS-Delayed}): algorihtm \ref{alg:trsm:recblasdelayed}.
\end{description}

We compare these three variants over finite fields with different cardinalities,
so as to make the parameters $t_\text{del}$ and $t_\text{update}$ vary as in
the following table:

\begin{center}
\begin{tabular}{|c||c|c|c|}
\hline
$p$ & $\lceil \log_2 p\rceil$ & $t_\text{del}$ & $t_\text{update}$\\
\hline
5   &   3        &   23            &  2\,147\,483\,642\\
1\,048\,583 & 20     &   2             & 8190 \\
8\,388\,617 & 23     &   2             & 126 \\
\hline
\end{tabular}
\end{center}

\begin{figure}[htbp]\begin{center}
\includegraphics[width=0.692\textwidth]{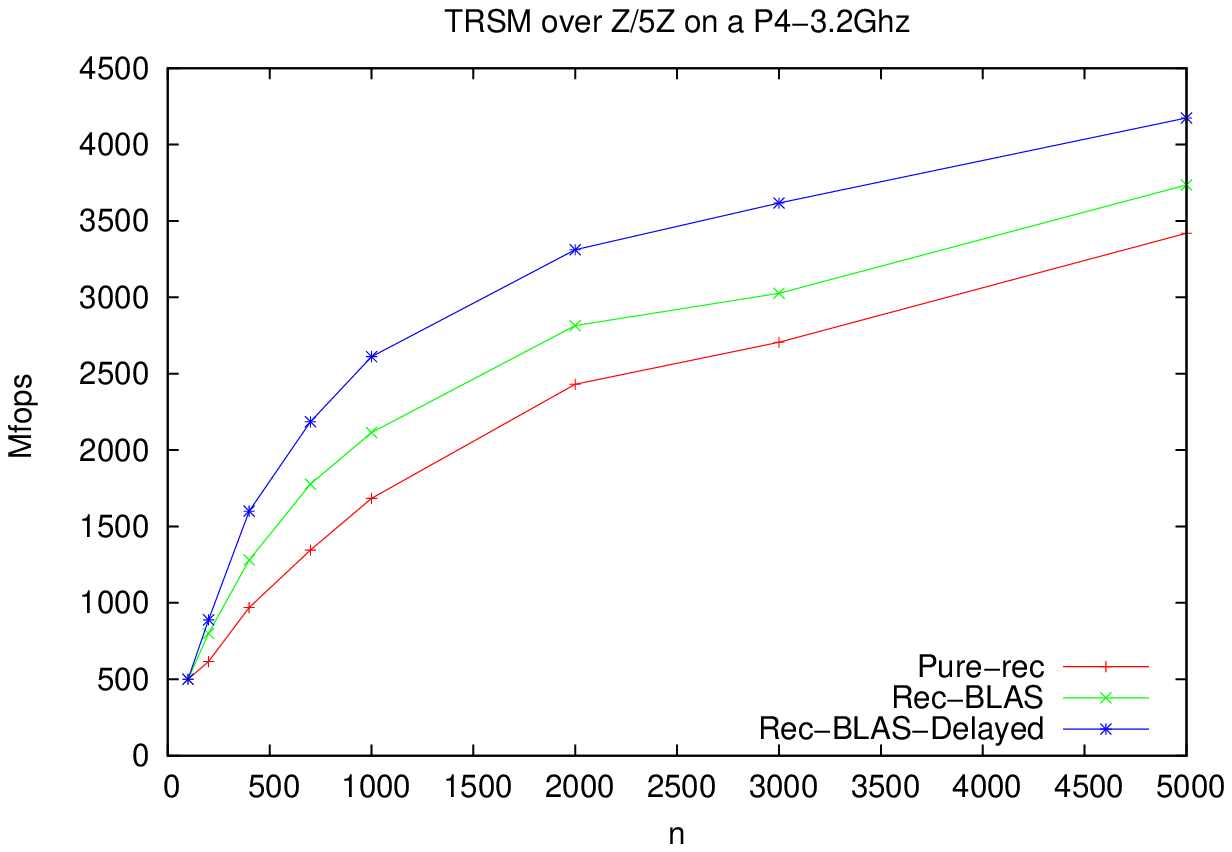}
\includegraphics[width=0.692\textwidth]{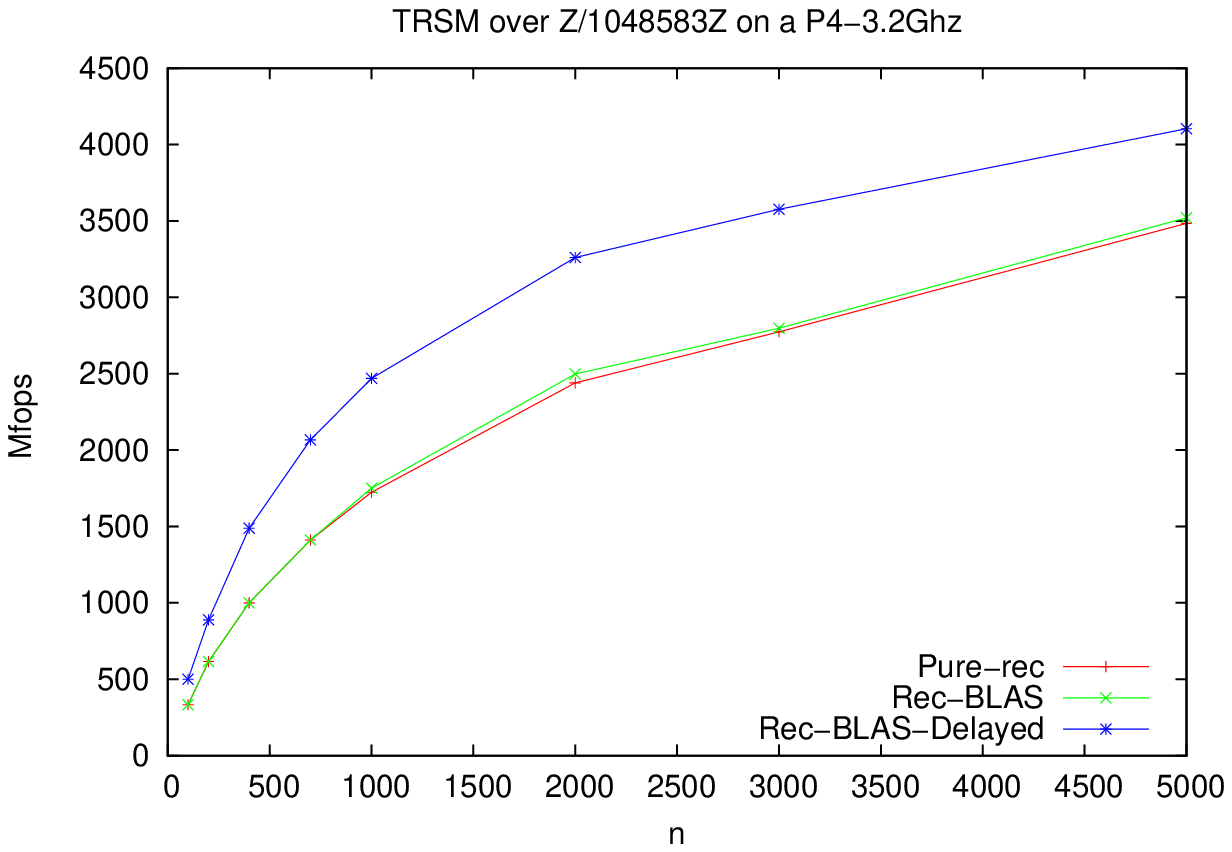}
\includegraphics[width=0.692\textwidth]{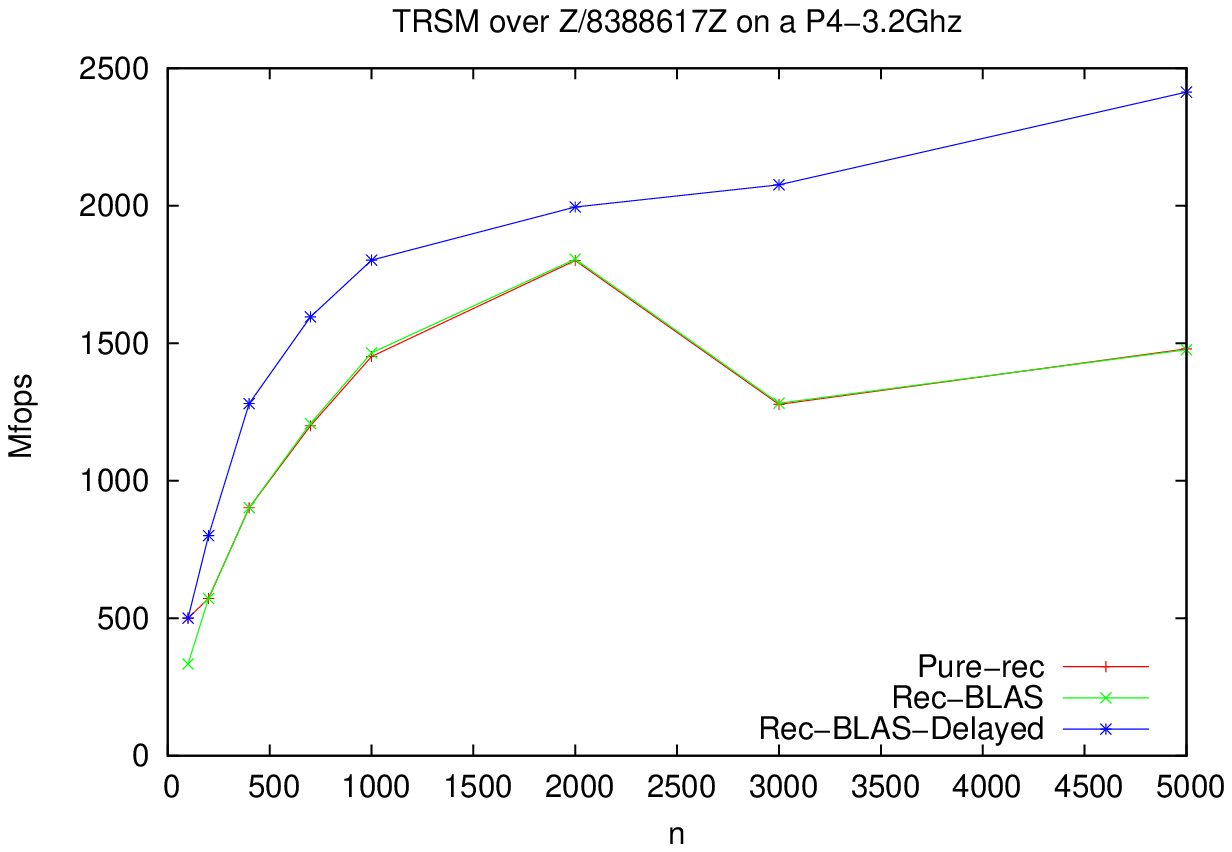}
\caption{Comparison of the \trsm variants for $p=5,1\,048\,583,8\,388\,617$,
 on a Pentium4\--3,2Ghz\--1Go}
\label{fig:trsm:compvar}
\end{center}\end{figure}

In the experiments of figure \ref{fig:trsm:compvar}, the matrix $B$ is 
square ($m=n$).
One can first notice the gain provided by the use of the first cascade with the
delayed \dtrsm routine by comparing the curves \texttt{rec-BLAS} and
\texttt{pure-rec} for $p=5$. This advantage shrinks when the characteristic gets larger,
 since $t_\text{del}=2$ for $p=1\,048\,583$ or $p=8\,388\,61$. 

Now the introduction of the coarse grain splitting, delaying the reductions in
the update phase improves by up to 500 Mfops the computation speed.
This gain is similar for $p=5$ and  $p=1\,048\,583$ since in both
cases  $n<t_\text{update}$ and there is therefore no modular reduction between the
matrix multiplications.

Lastly for $p=8\,388\,617$, the speed drops down since more reductions are required.
 The variants \texttt{pure-rec} and \texttt{rec-BLAS} are penalized by their 
dichotomic splitting, creating too many modular reductions after each matrix
multiplication. Now \texttt{rec-BLAS-delayed} has the best efficiency since the double
cascade structure minimizes the number of reductions.

We now give a comparison of this implementation with the equivalent routine of the original BLAS \dtrsm.
As for matrix multiplication in section \ref{ssec:fgemm-perf}, we compare the routines according to 
two different BLAS implementations (i.e. ATLAS and GOTO) and 
two different architectures. Nevertheless, we do not present the
results with ATLAS on Xeon architecture due to the surprisingly poor efficiency
of ATLAS \dtrsm during our tests.
In the following, \ftrsm denotes the \trsm routine over $16$-bits prime field (i.e. $\Z_{65521}$) 
using the \texttt{ZpZ-double} implementation.


\begin{table}[htbp]\begin{center}
\begin{tabular}{|cc|c||r|r|r|r|r|r|r|r|r|}
\cline{3-11}
\multicolumn{2}{c|}{} & $n$     
& {\em 1000}  & {\em 2000}  & {\em 3000}  & {\em 5000} & {\em 7000}  & {\em 8000} & {\em 9000} & {\em 10000} \\
\cline{3-11}
\multicolumn{11}{c}{}\\[-0.1cm]
\hline
&ATLAS & ftrsm & $0.37$s    &  $1.93$s    & $5.73$s    &   $23.63$s &  $62.50$s   & $91.67$s  & $121.84$s  & $166.74$s \\ 
\hline
\multicolumn{11}{c}{}\\[-0.2cm]
\hline

& & ftrsm  
&  $0.25$s    & $1.66$s    &  $5.08$s &
$21.47$s   & $55.95$s  & $80.77$s  & $111.57$s & $150.81$s \\ 
\cline{3-11}
& & dtrsm 
& $0.17$s    &  $1.35$s    & $4.50$s   &
$20.64$s &   $56.19$s   & $83.85$s  & $119.18$s & $163.33$s \\ 
\cline{3-11} \\[-.3cm]
\cline{3-11}
 & \begin{rotate}{90}\scriptsize GOTO \end{rotate} & {\bf
   $\frac{ftrsm}{dtrsm}$}  
& \bf 1.47 & \bf 1.23  &  \bf 1.13 &  \bf 1.04 & \bf 1.00 & \bf 0.96 & \bf 0.94 & \bf 0.92 \\   
\hline

\end{tabular}
\caption{Timings of triangular solver with matrix hand side on a Xeon,
  3.6GHz}\label{tab:trsm-p4}


\begin{tabular}{|cc|c||r|r|r|r|r|r|r|r|}
\multicolumn{11}{c}{}\\
\cline{3-11}
\multicolumn{2}{c|}{} & $n$          & {\em 1000}  & {\em 2000}  & {\em 3000}  & {\em 5000} & {\em 7000}  & {\em 8000} & {\em 9000} & {\em 10000}\\
\cline{3-11}
\multicolumn{11}{c}{}\\[-0.2cm]
\hline
& & ftrsm & $0.34$s & $2.28$s & $7.11$s & $30.26$s & $77.43$s & $112.01$s & $158.00$s & $214.31$s  \\
\cline{3-11}
& & dtrsm & $0.26$s & $1.95$s & $6.37$s & $28.60$s & $76.44$s & $113.78$s & $161.19$s & $219.31$s  \\
\cline{3-11} \\[-.3cm]
\cline{3-11}
& \begin{rotate}{90}\scriptsize ATLAS \end{rotate} & {\bf $\frac{ftrsm}{dtrsm}$}  & \bf 1.31 & \bf 1.17 & \bf 1.12 & \bf 1.06 & \bf 1.01 & \bf 0.98 & \bf 0.98 & \bf 0.98 \\
\hline
\multicolumn{11}{c}{}\\[-0.2cm]
\hline

& & ftrsm  & $0.30$s & $2.00$s & $6.23$s & $26.67$s & $68.22$s & $104.32$s & $137.96$s & $192.37$s  \\
\cline{3-11}
& & dtrsm  & $0.21$s & $1.61$s & $5.36$s & $24.59$s & $67.35$s & $100.42$s & $142.43$s & $195.79$s  \\
\cline{3-11} \\[-.3cm]
\cline{3-11}
 & \begin{rotate}{90}\scriptsize GOTO \end{rotate} & {\bf $\frac{ftrsm}{dtrsm}$}  & \bf 1.43 & \bf 1.24 & \bf 1.16 & \bf 1.08 & \bf 1.01 & \bf 1.04 & \bf 0.97 & \bf 0.98 \\
\hline

\end{tabular}
\caption{Timings of triangular solver with matrix hand side on Itanium2, 1.3GHz}\label{tab:trsm-ia64}
\end{center}
\end{table}

Tables \ref{tab:trsm-p4} and \ref{tab:trsm-ia64} show that our
implementation of exact {\trsm} solving is not far from numerical
performances.
Moreover, on our Xeon architecture, with GOTO BLAS, we are able to
achieve even better performances than numerical solving for matrices
of dimension greater than $7\,000$.

\begin{figure}[hbtp]
\begin{center}
\includegraphics[width=0.55\textwidth,angle=-90]{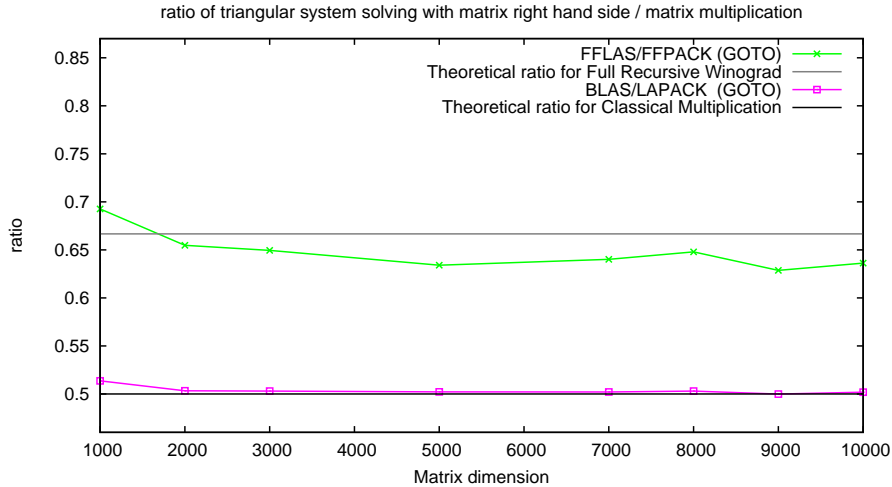}
\end{center}
\caption{Comparing  triangular system solving with matrix
  multiplication on a Xeon,
  3.6GHz} \label{fig:trsm-ratio}
\end{figure}

The good performance of our implementation is mostly achieved with 
the efficient reduction to fast matrix multiplication and the double
cascade structure.
Figure~\ref{fig:trsm-ratio} shows the ratio of the computation time of
our \trsm compared with matrix multiplication routine.
According to lemma \ref{lem:trsm}, this ratio is $1/2$ with $\omega=3$ 
and $2/3$ with $\omega = \log_2 7$.
In practice, our implementation only performs a few recursive calls of 
Winograd's algorithm, and the ratio appears to be between $0.5$ and $0.666$ as 
soon as the dimension is large enough, showing the good efficiency of the reduction to 
matrix multiplication.

%
%
%

%% file: ffpack.tex
\section{Finite Field Matrix Factorizations}\label{sec:pack}

We now come to one of the major interest of linear algebra over finite field:  matrix multiplication based algorithms.
The classical block Gaussian elimination is one of the most common
algorithm to achieve a reduction to matrix multiplication \cite{Strassen:1969:GENO}.
Nevertheless, our main concern here is the singularity of the matrices since we want to derive efficient algorithms
 for most problems (e.g. rank or nullspace).
One approach there is then to use a triangular form of the input matrix.
Hence, matrix triangularization algorithm plays a central role for this approach.
In this section we focus on practical implementations of triangularization in order to efficiently deal with rank profile, 
unbalanced dimensions, memory management, recursive thresholds, etc.
In particular we demonstrate the efficiency of matrix multiplication reduction  in practice for many linear algebra problems.

\input{triangul}

%
\input{use}
%
%

%% file: triangul.tex
\subsection{Triangularizations} \label{sec:triang}

The classical block $LDU$ or $LUP$ factorizations (see \cite{Aho:1974:DACA}) can not be used due to their restriction to non-singular case.
Instead one would rather use the \lqup\ factorization of \cite{Ibarra:1982:LSP}. 
We here propose a fully in-place variant and analyze its behaviour.

\input{lqup}

\subsection{Performance and comparison with numerical routines}

Fast matrix multiplication routine of section \ref{ssec:winograd} allowed us to speed up matrix multiplication as well as 
triangular system solving. These improvements are of great interest since they directly improve efficiency of triangularization.
We now compare our exact triangularization over finite field with numerical triangularization provided within LAPACK library \cite{1999:lapack}.
In particular, we use an optimized version of this library provided by ATLAS software in which we use two different BLAS kernel: ATLAS and GOTO.

\begin{table}[htb]\begin{center}
\begin{tabular}{|cc|c||r|r|r|r|r|r|r|r|r|}
\cline{3-11}
\multicolumn{2}{c|}{} & $n$        
& {\em 1000}  & {\em
  2000}  & {\em 3000}  & {\em 5000} & {\em 7000}  & {\em 8000} & {\em
  9000} & {\em 10000} \\
\cline{3-11}
\multicolumn{11}{c}{}\\[-0.2cm]
\hline
& & lqup 
& $0.32$s    &  $1.84$s    & $4.89$s    &
$19.34$s &  $48.94$s   & $73.86$s  & $97.50$s & $131.11$s \\ 
\cline{3-11}
& & dgetrf 
& $0.17$s    &  $1.19$s    & $3.83$s   &
$16.90$s &  $45.32$s   & $67.44$s  & $94.83$s & $130.15$s \\ 
\cline{3-11} \\[-.3cm]
\cline{3-11}
& \begin{rotate}{90}\scriptsize ATLAS \end{rotate} & {\bf
  $\frac{lqup}{dgetrf}$}  
& \bf 1.88  & \bf 1.55 & \bf 1.28  &  \bf 1.14 &  \bf 1.08 & \bf 1.10 & \bf 1.03 & \bf 1.01  \\   
\hline
\multicolumn{11}{c}{}\\[-0.2cm]
\hline
& & lqup 
& $0.25$s    &  $1.52$s    & $4.47$s    &
$17.93$s &  $44.54$s   & $67.88$s  & $89.63$s & $119.65$s \\ 
\cline{3-11}
& & dgetrf 
& $0.15$s    &  $1.03$s    & $3.33$s   &
$14.84$s &  $39.58$s   & $58.61$s  & $82.89$s & $113.47$s \\ 
\cline{3-11} \\[-.3cm]
\cline{3-11}
& \begin{rotate}{90}\scriptsize GOTO \end{rotate} & {\bf
  $\frac{lqup}{dgetrf}$}  
& \bf 1.67  & \bf 1.48 & \bf 1.34  &  \bf 1.21  &  \bf 1.13 & \bf  1.16 & \bf 1.08  & \bf 1.05  \\   
\hline

\end{tabular}
\caption{Performance of matrix triangularization (for $\pF{65521}$ and
  floats) on a Xeon, 3.6GHz}\label{tab:lqup-p4}


\begin{tabular}{|cc|c||r|r|r|r|r|r|r|r|}
\multicolumn{11}{c}{}\\
\cline{3-11}
\multicolumn{2}{c|}{} & $n$          & {\em 1000}  & {\em 2000}  & {\em 3000}  & {\em 5000} & {\em 7000}  & {\em 8000} & {\em 9000} & {\em 10000}\\
\cline{3-11}
\multicolumn{11}{c}{}\\[-0.2cm]
\hline
& & lqup & $0.38$s & $2.20$s & $6.36$s & $25.22$s & $61.64$s & $89.74$s & $127.43$s & $163.68$s  \\
\cline{3-11}
& & dgetrf & $0.20$s & $1.47$s & $4.61$s & $20.26$s & $53.57$s & $79.37$s & $111.66$s & $152.42$s  \\
\cline{3-11} \\[-.3cm]
\cline{3-11}
& \begin{rotate}{90}\scriptsize ATLAS \end{rotate} & {\bf $\frac{lqup}{dgetrf}$}  & \bf 1.85 & \bf 1.50 & \bf 1.38 & \bf 1.25 & \bf 1.15 & \bf 1.13 & \bf 1.14 & \bf 1.07 \\
\hline
\multicolumn{11}{c}{}\\[-0.2cm]
\hline
& & lqup & $0.34$s & $2.00$s & $5.81$s & $23.11$s & $56.80$s & $83.90$s & $113.66$s & $150.82$s  \\
\cline{3-11}
& & dgetrf & $0.16$s & $1.17$s & $3.80$s & $17.07$s & $46.18$s & $69.00$s & $97.56$s & $134.01$s  \\
\cline{3-11} \\[-.3cm]
\cline{3-11}
& \begin{rotate}{90}\scriptsize GOTO \end{rotate} & {\bf $\frac{lqup}{dgetrf}$}  & \bf 2.21 & \bf 1.72 & \bf 1.53 & \bf 1.35 & \bf 1.23 & \bf 1.22 & \bf 1.16 & \bf 1.13 \\
\hline
\end{tabular}
\caption{Performance of matrix triangularization (for $\pF{65521}$ and
  floats) on Itanium2-1.3GHz}\label{tab:lqup-ia64}
\end{center}
\end{table}

Tables  \ref{tab:lqup-p4} and \ref{tab:lqup-ia64} show efficiency obtained  with our exact triangularization based on fast matrix multiplication
and the one obtained with numerical computation. There, 
``dgetrf'' computes a floating point LU factorization of a general 
$m \times n$  matrix  using partial pivoting with row interchanges.
Exact computation is done in the prime field of integers modulo $65521$.
We are now mostly able to reach the speed of numerical computations.
More precisely, we are able to compute the triangularization of a $10\,000 \times 10\,000$ matrix over a finite field in about $2$ minutes on a  Xeon 3.6GHz architecture.
This is only $5\%$ slower than the best numerical computation.

We could have expected that our speed would have been even better than numerical approach since we take advantage of Strassen-Winograd's
multiplication while numerical computations are not.
However, in practice we do not fully benefit from fast matrix multiplication since we work at most with matrices of half dimension of the input matrix due to the recursive structure of the algorithm.
Then, the number of Winograd calls is at least one less than within matrix multiplication routines.
In our tests, it appears that we only use 3 calls on our Xeon architecture and 1 call on the Itanium2 architecture according to matrix multiplication threshold.
This explains the better performance on the Xeon compared to numerical
routines than the Itanium2 architecture. 

%
%
%
%
%
%
%
Note also that in order to take even more into account data locality one can develop a version of \lqup\ where blocks
 are maintained as square as possible. 
Indeed, as soon as the RAM is full, data locality becomes more important than memory saves.
The \tu\ method \cite{jgd:2002:PComp} addresses this issue.
A first implementation of \tu\ has been studied in \cite[\S4.5]{jgd:2004:ffpack} and it reveals
to be the fastest for large matrices, despite its bigger memory demand \cite[Figure 6]{jgd:2004:ffpack}.
This is advocating further uses of recursive blocked data formats and of more recursive levels of \tu.

\subsection{Comparison with the multiplication}
\begin{figure}[hbtp]
\begin{center}
\includegraphics[width=0.55\textwidth,angle=-90]{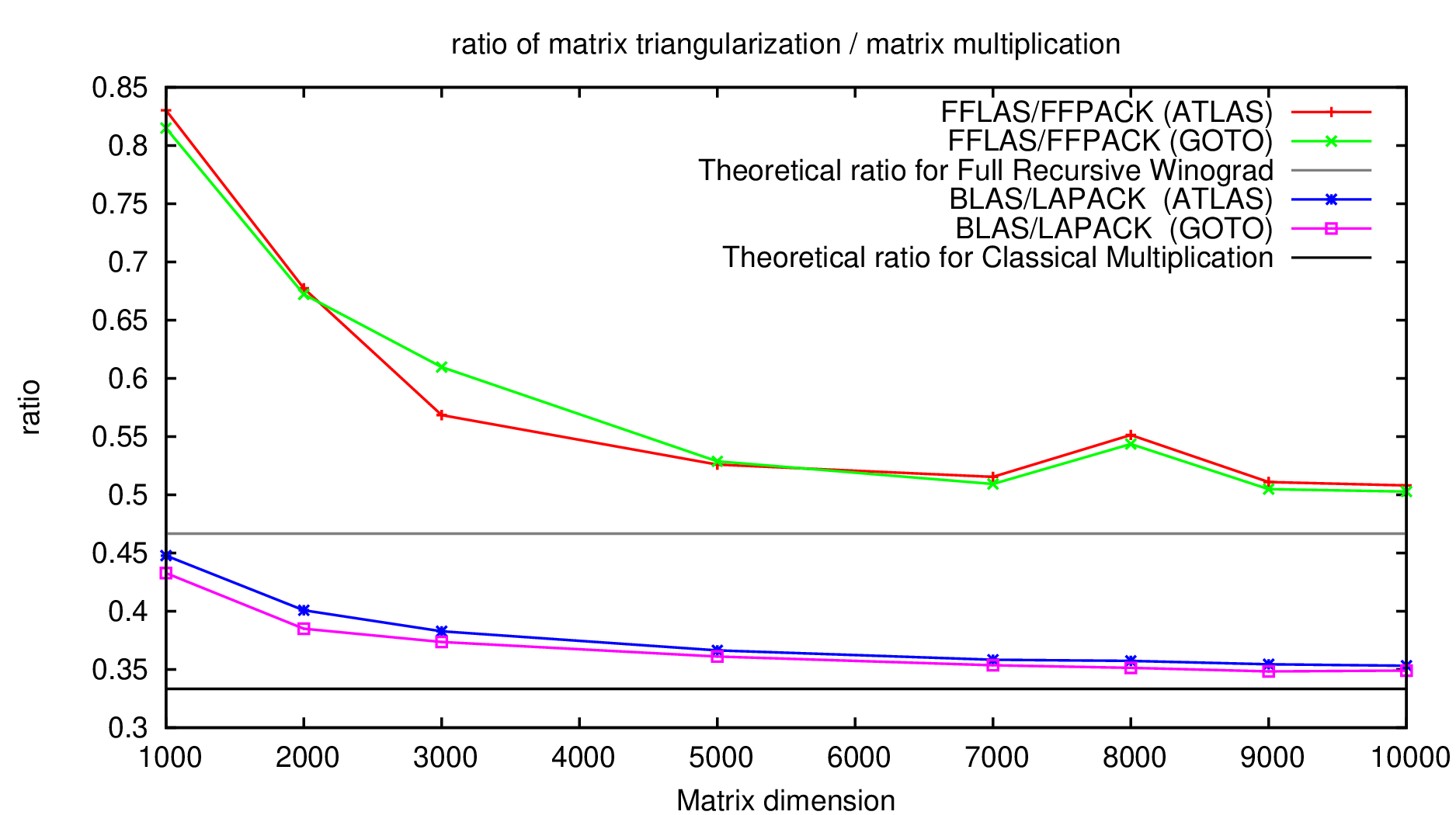}
\end{center}
\caption{Comparing  matrix triangularization  with matrix
  multiplication on a Xeon,
  3.6GHz}\label{fig:ratio-lqup}
\end{figure}

The \lqup\ factorization and the \trsm\ routines  reduce to matrix multiplication as we have seen in the previous sections. 
Theoretically, as classic matrix multiplication requires $2n^3-n^2$ arithmetic operations, the factorization, requiring at most
$\frac{2}{3}n^3$ arithmetic operations, could be computed in about $\frac{1}{3}$ of the time. 
However, when Winograd fast matrix multiplication algorithm is used this ratio becomes $\frac{2}{5}$.
Figure  \ref{fig:ratio-lqup} shows that the experimental behavior of
the factorization 
is not very far from this theoretical ratio.

%
%
%
%

%% file: lqup.tex
The \lqup\  factorization is a generalization of the well known block \lup\ 
factorization for the singular case \cite{BunchHopcroft:1974}.
Let $A$ be a $m \times n$ matrix, we want to compute the quadruple
$<L,Q,U,P>$ such that $A=LQUP$.
The matrix {\tt L} is lower triangular, {\tt P} and {\tt Q} are permutation
matrices  and {\tt U} is a rank $r$ upper triangular matrix
with its $r$ first rows non-zero.

The algorithm with best known complexity computing this
factorization uses a divide and conquer approach and 
reduces to matrix multiplication \cite{Ibarra:1982:LSP}. 
Let us describe briefly the behavior of this algorithm.

\begin{figure}[htbp]\begin{center}
\includegraphics[width=0.34\columnwidth]{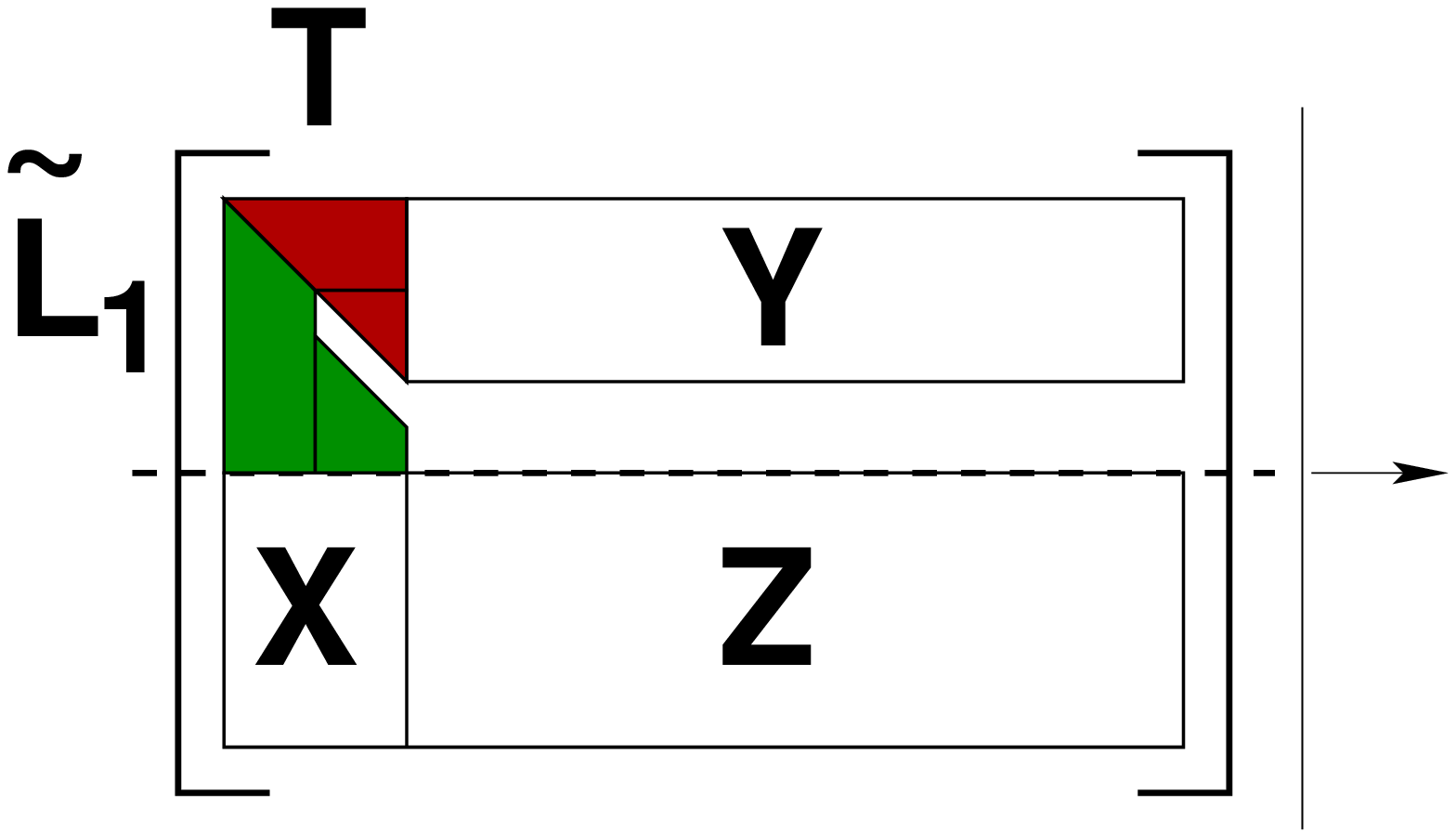}
\includegraphics[width=0.3\columnwidth]{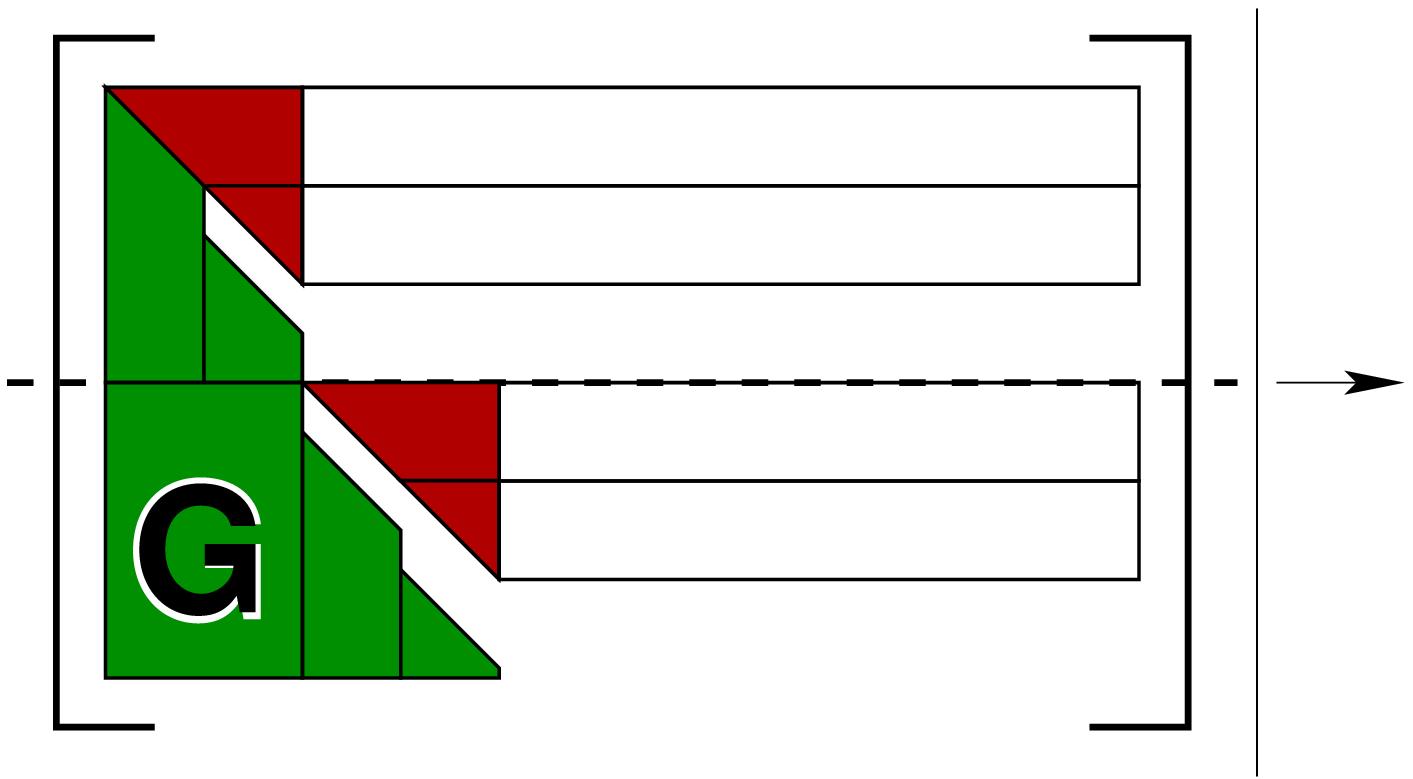}
\includegraphics[width=0.33\columnwidth]{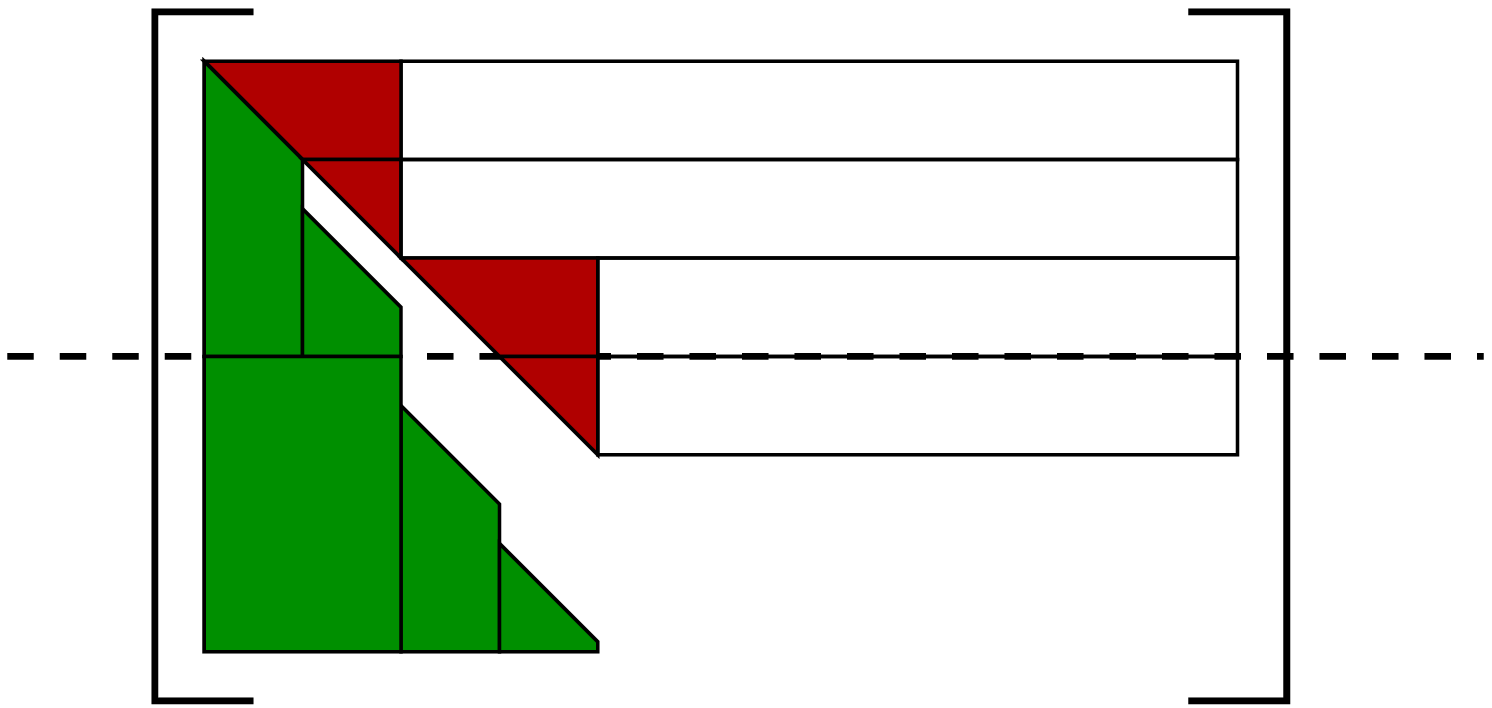}
\caption{Principle of the LQUP factorization}
\label{fig:lqup}
\vspace{-1ex}
\end{center}\end{figure}

The algorithm is recursive: first, it splits $A$ in halves and
performs a recursive call on the top half.
After some row permutations,
It thus gives the $T$, $Y$ and $L_1$ blocks of figure \ref{fig:lqup}, together with some row permutations stored in $Q$.
Then, after some column permutations ($[X Z]=[A_{21} A_{22}]P$), 
the algorithm computes $G$ such that $G T = X$ via {\tt trsm}, replaces
$X$ by zeroes and eventually updates $Z = Z - G Y$. 
The third step is a recursive call on $Z$, followed by an update of $Q$.
We let the  readers refer e.g. to \cite[(2.7c)]{Bini:1994:PMCFA} for
further details. 

 Furthermore, our implementation of \lqup\  also uses
the trick proposed in \cite[\S4.2]{jgd:2004:ffpack}, namely storing $L$ in its compressed form
$\tilde{L}$. 

This triangularization is thus fully in-place.

\begin{lem} The dominant term of the time complexity of algorithm \lqup
 with $m \leq n$ is 
%
$$\lqup(m,n)=\left( \lCeil \frac{n}{m} \rCeil
  \frac{1}{2^{\omega-1}-2}-\frac{1}{2^\omega-2}\right) \MM(m).$$
The latter is $nm^2-\frac{1}{3} m^3$ with classical multiplication.

\end{lem}

\begin{proof}
Lemma \ref{lem:mn} ensures that the cost is
$\GO(m^\omega+nm^{\omega-1})$. We thus just have to look for the
constant factors. Then we write 
$\lqup(m,n)=\alpha m^\omega+\beta
nm^{\omega-1}=\lqup(m/2,n)+\TRSM(m/2,r)+R(m/2,r,n-r)+\lqup(m/2,n-r)$, where
$r$ is the rank of the first $m/2$ rows. This
gives $\alpha m^\omega+\beta
nm^{\omega-1}= \alpha (m/2)^\omega+\beta
n(m/2)^{\omega-1}+\frac{1}{2^{\omega-1}-2}\lCeil \frac{m}{2r}\rCeil
\MM(r)+\lCeil\frac{m(n-r)}{2r^2}\rCeil \MM(r) + \alpha (m/2)^\omega+\beta
(n-r)(m/2)^{\omega-1}$.
With $m \leq n$, the latter is maximal for $r = m/2$, and
then, writing
$\MM(x)=C_\omega x^\omega$, we identify the coefficient on both sides:
$\beta=\frac{\beta}{2^{\omega-1}}+\frac{C_\omega}{2^{\omega-1}}+\frac{\beta}{2^{\omega-1}}$,
and $\alpha=2\frac{\alpha}{2^\omega}-\frac{\beta}{2^\omega}-C_\omega\frac{2^\omega-6}{2^\omega(2^\omega-4)}$.
Solving for $\beta$ and $\alpha$ gives the announced terms.
\end{proof}


%% file: use.tex
\section{Applications}\label{sec:use}
In this section, we use our matrix multiplication, matrix
factorization and matrix solvers as basic routines to perform other
linear algebra routines. 
For instance, from the two routines (i.e. \lqup\ and \trsm),
 one can also directly derive several other algorithms, e.g.:
\begin{itemize}
\item The {\bf rank} is
the number of non-zero rows in $U$.
\item The {\bf
  determinant} is the product of the diagonal elements of $U$ (stopping whenever a zero is encountered).
\end{itemize}

In the following, we first give
the theoretical complexities with explicit constant terms. These
constants depend on the kind of matrix multiplication used (fast or
classical). In order to validate our approach we then compare this
theoretical ratios to some experimental ones. 

\subsection{Nullspace basis}
Computing a right nullspace basis with the LQUP factorization is immediate on a  $m \times n$ full rank matrix,
where $m \leq n$: if $U=[U_1 U_2]$, the matrix $U_1^{-1} U_2$ completed with identity matrix yields a basis for the nullspace of $A$.

This requires 
$NS(m;n)=LQUP(m;n)+\TRSM(m;n-m)$.
which gives
\begin{equation}
NS(m;n) = (\lCeil\frac{n}{m} \rCeil \frac{2}{2^{\omega-1}-2} -\frac{1}{2^{\omega}-2} ) \MM(m)
\end{equation}
The latter is
$(m^2n-\frac{1}{3}m^3)+(n-m)m^2=2m^2n-\frac{4}{3}m^3$ with classical multiplication.
One can notice that computing a right nullspace of the transposed of the input matrix yields a left nullspace basis.

\subsection{Triangular multiplications}
\subsubsection{Triangular matrix multiplication}
To perform the multiplication of a triangular matrix by a dense
matrix via a block decomposition in halves, one requires four recursive calls
and two dense matrix-matrix multiplications. The cost is thus 
$TRMM(n) = 4 TRMM(n/2) + 2 \MM(n/2)$, solving for 
$TRMM(n)=\alpha \MM(n)$ yields 
\begin{equation} TRMM(n) = \frac{1}{2^{\omega-1}-2} \MM(n).\end{equation}
The latter is $n^3$ with classical multiplication.

\subsubsection{Upper-lower Triangular matrix multiplication}
The block multiplication of a  lower triangular matrix by an
upper
triangular 
matrix is 
\[ 
\begin{array}{ccccc}
\left[ \begin{array}{cc} A_1 & A_2 \\  & A_4
    \end{array} \right]  & \times &
\left[ \begin{array}{cc} B_1 &  \\ B_3 & B_4  \end{array} \right]&
= &
\left[ \begin{array}{cc} A_1 B_1 + A_2 B_3 & A_2 B_4
      \\ A_4 B_3 & A_4 B_4 \end{array} \right]
\end{array}
\]
The cost is thus 
$\UTLT(n) = 2 \UTLT(n/2) + 2 TRMM(n/2) + \MM(n/2)$, solving for 
$\UTLT(n)=\alpha \MM(n)$ yields 
\begin{equation} \UTLT(n) = \frac{2^{\omega}}{(2^{\omega}-4)(2^{\omega}-2)} \MM(n).\end{equation}
The latter is $\frac{2}{3}n^3$ with classical multiplication.

\subsubsection{Upper-Upper Triangular matrix multiplication}
Now the block version is even simpler (of course the lower lower
multiplication is similar):
\[ 
\begin{array}{ccccc}
\left[ \begin{array}{cc} A_1 & A_2 \\  & A_4
    \end{array} \right]  & \times &
\left[ \begin{array}{cc} B_1 & B_2 \\  & B_4  \end{array} \right]&
= &
\left[ \begin{array}{cc} A_1 B_1 & A_1 B_2 + A_2 B_4
      \\  & A_4 B_4 \end{array} \right]
\end{array}
\]
The cost is thus 
$\UTUT(n) = 2 \UTUT(n/2) + 2 TRMM(n/2)$, which yields 
\begin{equation} \UTUT(n) = \frac{4}{(2^{\omega}-4)(2^{\omega}-2)} \MM(n).\end{equation}
The latter is $\frac{1}{3}n^3$ with classical multiplication.

\subsection{Squaring}
\subsubsection{$A \times A^T$}\label{ssec:aat}
Suppose we want to compute $A$ times its transpose, even with a diagonal in the middle.
The block version is 
{\small
\[ 
\left[ \begin{array}{cc} A_1 & A_2 \\ A_3 & A_4
    \end{array} \right]   \times 
\left[ \begin{array}{cc} D_1 & \\ & D_4
    \end{array} \right]  \times 
\left[ \begin{array}{cc} A_1^T & A_3^T \\ A_2^T & A_4^T  \end{array} \right]
= 
\left[ \begin{array}{cc} A_1 D_1 A_1^T + A_2 D_4 A_2^T & A_1 D_1 A_3^T
      + A_2 D_4 A_4^T \\ A_3 D_1 A_1^T + A_4 D_4 A_2^T & A_3 D_1 A_3^T
      + A_4 D_4 A_4^T \end{array} \right]
\]}

Since $A D A^T$ is symmetric, the lower left and upper right are just
transpose of one another. The other corners (upper left and lower right)
are computed
via recursive calls.
Thus
the arithmetic cost of this special product is 
$AAT(n) = 4 AAT(n/2) + 2\MM(n/2) + 3\ADD(n/2) + 2 (n/2)^2$

Ignoring the cost of the three additions and the diagonal multiplications, 
this yields 
\begin{equation} AAT(n) = \frac{2}{2^{\omega}-4} \MM(n).\end{equation}
The latter is $n^3$ with classical multiplication.
One can note that when $A$ is rectangular with $m \leq n$ the cost extends to
\begin{equation}
AAT(m;n) = \lCeil\frac{n}{m}\rCeil\frac{2}{2^{\omega}-4} \MM(m).
\end{equation}

\subsubsection{Symmetric case}
When $A$ is already symmetric, and if the diagonal is unitary,
the constant factor decreases.
Indeed, in this case $A_2=A_3^T$ and then one of the four recursive
calls is saved. Also one of the remaining three recursive calls is a
call
to a non symmetric $A A^T$. Therefore the cost is now:
$SymAAT(n) = 2 SymAAT(n/2) + AAT(n/2) + 2\MM(n/2)$, once again ignoring $n^2$.
This yields
\begin{equation} SymAAT(n) = \frac{2(2^{\omega}-3)}{(2^{\omega}-4)(2^{\omega}-2)} \MM(n).\end{equation}
The latter is $\frac{5}{6} n^3$ with classical multiplication.
\subsubsection{Triangular case}
We here view the explicit computation of $L^T D L$ for instance as a special case of upper-lower 
triangular matrix multiplication, but where both matrices are symmetric of one another. We also show that we can add an extra diagonal factor in the middle at a negligible cost.
Consider then 
\[ 
\begin{array}{ccccccc}
\left[ \begin{array}{cc} L_1 & \\ L_3 & L_4
    \end{array} \right]  & \times & 
\left[ \begin{array}{cc} D_1 & \\ & D_4
    \end{array} \right]  & \times &
\left[ \begin{array}{cc} L_1^T & L_3^T \\ & L_4^T  \end{array} \right]&
= &
\left[ \begin{array}{cc} L_1 D_1 L_1^T & L_1 D_1 L_3^T
       \\ L_3 D_1 L_1^T & 
      L_3 D_1 L_3^T + L_4 D_4 L_4^T \end{array} \right]
\end{array}
\]

Thus it requires two recursive calls, a call to AAT (with a diagonal in the middle) only one call to TRMM 
as both lower-left and upper-right corners are transpose of one another.
This yields 
\begin{equation} \LTL(n) = \frac{4}{(2^{\omega}-4)(2^{\omega}-2)} \MM(n).\end{equation}
The latter is $\frac{1}{3} n^3$ with classical multiplication.

\subsection{Symmetric factorization}
For the sake of simplicity, 
we here consider the $LU$ factorization of a generic rank profile symmetric 
$n \times n$ matrix $A$. 
We could describe how to perform this decomposition with the
permutation and the possible rank deficiency in the blocks,
but we here only
analyze the cost of such a $LDL^T$ factorization.
The idea is that one can recursively decompose
$A=\left[\begin{array}{cc}A_1&A_2\\A_2^T&A_4\end{array}\right]=
\left[\begin{array}{cc}L_1&\\G&L_2\end{array}\right]
\times 
\left[\begin{array}{cc}D_1&\\&D_2\end{array}\right]
\times 
\left[\begin{array}{cc}L_1^T&G^T\\&L_2^T\end{array}\right]
$.
Well, this requires a recursive call to compute $L_1$ and $D_1$ ; a TRSM
to compute $G$ such that $L_1 D_1 G^T = A_2$ ; an AAT to compute $GD_1G^T$
and
a recursive call to compute $L_2D_2L_2^T = A_4-GD_1G^T$.
The cost is thus $LDLT(n)=2LDLT(n/2)+\TRSM(n/2)+AAT(n/2)$, which yields
\begin{equation} LDLT(n) = \frac{4}{(2^{\omega}-4)(2^{\omega}-2)} \MM(n).\end{equation} 
The latter is $\frac{1}{3} n^3$ with classical multiplication.
%
\subsection{Matrix inverse}

\subsubsection{Triangular matrix inverse}\label{sssec:trinverse}
To invert a triangular matrix via a block decomposition, 
one requires two recursive calls and two triangular matrix
multiplications.
\[ 
\begin{array}{ccc}
\left[ \begin{array}{cc} A_1 & A_2\\  & A_4
    \end{array} \right]^{-1}  &  = &
  
\left[ \begin{array}{cc} A_1^{-1} & -A_1^{-1}A_2A_4^{-1} \\ & A_4^{-1}  \end{array} \right]
\end{array}
\]

The cost is thus $\INVT(n)=2\INVT(n/2)+2TRMM(n/2)$ which yields
\begin{equation} \INVT(n) = \frac{2}{2^{\omega}-2} TRMM(n) = \frac{4}{(2^{\omega}-4)(2^{\omega}-2)} \MM(n). \end{equation}
The latter is $\frac{1}{3} n^3$ with classical multiplication.

\subsubsection{Matrix inverse}\label{sssec:inverse}
To invert a dense matrix, one needs to compute an $LQUP$
decomposition, then to invert $L$ and permute it with $Q^{-1}$.
A TRSM is then required to solve $U X = Q^{-1}L^{-1}$. Applying
$P^{-1}$ to X yields the inverse.
The cost is then $INV(n)=LQUP(n)+\INVT(n)+\TRSM(n)$.
This gives \begin{equation} INV(n) = \frac{3\times 2^\omega}{(2^\omega-4)(2^\omega-2)} \MM(n).\end{equation}
The latter is $INV(n)= 2n^3$ with classical multiplication.

\subsubsection{Symmetric inverse}
If $A$ is symmetric, one can decompose it into a $LDL^T$ factorization instead of the $LU$.
Therefore, its inverse is then only one $INVT$ for both $L^{-1}$ and $L^{-T}$ followed by an $LTL$.
The cost is then 
$SymINV(n) =  LDLT(n)+\INVT(n)+\LTL(n)$
which yields 
\begin{equation}
SymINV(n) = \frac{12}{(2^\omega-2)(2^\omega-4)} \MM(n).
\end{equation}
The latter is $SymINV(n) = n^3$ with classical multiplication.

\subsubsection{Full-rank Moore-Penrose pseudo-inverse}
$A$ is a rectangular full rank $m
\times n$ matrix. 
We suppose, without loss of genericity, that $m \leq n$.
The Moore-Penrose inverse of $A$ is thus $A^{\dagger}=A^T (A A^T)^{-1}$,
see e.g. \cite{Saunders:2001:BBLS} and references therein.
Computing the Moore-Penrose inverse is then just a $LDL^T$ decomposition of the symmetric matrix $AA^T$, followed by two rectangular system solvings:
$$MPINV(m;n) = AAT(m;n) + LDLT(m) + 2 \TRSM(m;n).$$
The cost is then 
\begin{equation}
MPINV(m;n) = \left( \lCeil \frac{n}{m}\rCeil \frac{6}{2^\omega -4} + \frac{4}{(2^\omega -2)(2^\omega -4)}   \right) \MM(m)
\end{equation}
The latter is $3 m^2n + \frac{1}{3}m^3$ with classical
multiplication. This correspond e.g. to the normal equations numerical
resolution \cite[algorithm 5.3.1]{Golub:1996:MatrixC}.

\subsubsection{Rank deficient Moore-Penrose pseudo-inverse}
In this case, one needs to compute a full-rank decomposition of $A$. 
This is done by performing the $LQUP$
decomposition of $A$ and if $A$ is of rank $r$, selecting the first $r$
columns of $L$ (call them $L_r=\left[\begin{array}{c}L_1\\G\end{array}\right]$) and the first $r$ rows $U$ (call them
$U_r=[U_1|Y]$), forgetting the permutation
$P$. We have $A=L_r U_r$ and we modify the formula
\cite[(7)]{Noble:1966:MCG} as follows:
\begin{equation}
A^{\dagger}=\left[\begin{array}{c}I\\Y^TU_1^{-T}\end{array}\right]\left((L_1+L_1^{-T}G^TG)(U_1+YY^TU_1^{-1})\right)^{-1}[I|L_1^{-T}G^T].
\end{equation}
We note $W=(L_1+L_1^{-T}G^TG)(U_1+YY^TU_1^{-1})$.
We compute $W$ by two squarings, two TRSM and a classical matrix
multiplication.
We perform a reversed LU decomposition on $W$ to get $W=U_wL_w$.
Now we compute $L_1^TU_w$ and $L_wU_1^T$ by upper-upper triangular
multiplication and $H=(L_1^TU_w)^{-1}G^T$ and $Z=Y^T(L_wU_1^T)^{-1}$
by two TRSM.
Now, $A^{\dagger}=\left[\begin{array}{cc}W^{-1}&L_w^{-1}H\\ZU_w^{-1}&ZH\end{array}\right]$.
$W^{-1}$ is two triangular inverses and an upper lower product.
$ZH$ is a rectangular multiplication and the last two blocks are
obtained by two triangular solvings.
\begin{multline}
MPINV_r(m;n) =
\LQUP(m;n)+AAT(r;m-r)
+AAT(r;n-r)+3\TRSM(r,m-r)\\
+3\TRSM(r,n-r)+\MM(r)
+\LQUP(r)+2\UTUT(r)+2\INVT(r)+\UTLT(r)\\
+R(n-r;r;m-r)
\end{multline}
The latter is $2rmn+2r^2m+2r^2n+m^2n-\frac{1}{3}m^3-\frac{4}{3}r^3$ with
classical multiplication.
To get an idea, numerical computations based on the Cholesky
factorization of $AA^T$ presented in
\cite{Courrieu:2005:FMP} as faster than SVD or QR or iterative methods
would require $3m^2n+2r^2m+3r^3$ flops.
\subsubsection{Performances and comparisons with numerical routines}
As for triangular system solving and matrix triangularization, we now compare performances of matrix inversion for triangular and dense matrices
with numerical computation and with matrix multiplication.
Our comparison with numerical computation is still based on LAPACK library with two different BLAS kernel (i.e. ATLAS and GOTO).
We do not present the result of triangular matrix inversion over our Xeon architecture according to the bad behavior of ``dtrsm'' function which is the 
main routine used by LAPACK for triangular matrix inversion.
Our base field is the prime field of integers modulo $65521$ using a {\tt Zpz-double} representation and we use fast matrix multiplication of section \ref{ssec:winograd}.
%
%
%
%
%
%
\begin{table}[htbp]\begin{center}
\begin{tabular}{|cc|c||r|r|r|r|r|r|r|r|r|}
\cline{3-11}
\multicolumn{2}{c|}{} & $n$       
& {\em 1000}  & {\em 2000}  & {\em 3000}  & {\em 5000} & {\em 7000}  & {\em 8000} & {\em 9000}& {\em 10000}\\
\cline{3-11}
\multicolumn{11}{c}{}\\[-0.1cm]
\hline
&\scriptsize ATLAS & tri. inv & $0.11$s    &  $0.70$s    & $2.17$s    &   $9.21$s &  $24.21$s   & $35.53$s  & $49.95$s  & $68.26$s \\ 
\hline
\multicolumn{11}{c}{}\\[-0.2cm]
\hline
& & tri. inv 
& $0.10$s    &  $0.62$s    & $1.90$s    &
$8.00$s &  $20.97$s   & $30.77$s  & $43.38$s & $58.98$\\ 
\cline{3-11}
& & dtrtri 
&  $0.18$s  &  $1.04$s    & $2.90$s   &
$10.97$s &  $26.85$s   & $38.57$s  & $52.93$s & $70.95$s \\ 
\cline{3-11} \\[-.3cm]
\cline{3-11}
& \begin{rotate}{90}\scriptsize GOTO \end{rotate} & {\bf
  $\frac{tri. inv}{dtrtri}$}  
& \bf 0.56  & \bf 0.60  & \bf 0.66   &  \bf 0.73  &  \bf 0.78  & \bf 0.80  & \bf 0.82 & \bf 0.83  \\   
\hline

\end{tabular}
\caption{Timings of triangular matrix inversion on a Xeon, 3.6GHz}\label{tab:trinv-p4}


\begin{tabular}{|cc|c||r|r|r|r|r|r|r|r|}
\multicolumn{11}{c}{}\\
\cline{3-11}
\multicolumn{2}{c|}{} & $n$          & {\em 1000}  & {\em 2000}  & {\em 3000}  & {\em 5000} & {\em 7000}  & {\em 8000} & {\em 9000} &{\em 10000}\\
\cline{3-11}
\multicolumn{11}{c}{}\\[-0.2cm]
\hline
& & tri. inv & $0.19$s & $1.03$s & $3.02$s & $11.91$s & $31.71$s & $44.43$s & $61.37$s & $82.55$s  \\
\cline{3-11}
& & dtrtri& $0.08$s & $0.58$s & $2.55$s & $11.39$s & $30.50$s & $44.52$s & $63.34$s & $85.19$s  \\
\cline{3-11} \\[-.3cm]
\cline{3-11}
& \begin{rotate}{90}\scriptsize ATLAS \end{rotate} & {\bf $\frac{tri. inv}{dtrtri}$}  & \bf 2.25 & \bf 1.77 & \bf 1.18 & \bf 1.05 & \bf 1.04 & \bf 1.00 & \bf 0.97 & \bf 0.97 \\
\hline
\multicolumn{11}{c}{}\\[-0.2cm]
\hline
& & tri. inv & $0.15$s & $0.85$s & $2.47$s & $10.10$s & $26.10$s & $38.29$s & $53.65$s & $72.74$s  \\
\cline{3-11}
& & dtrtri & $0.08$s & $0.61$s & $1.96$s & $8.77$s & $23.68$s & $35.73$s & $49.84$s & $69.10$s  \\
\cline{3-11} \\[-.3cm]
\cline{3-11}
& \begin{rotate}{90}\scriptsize GOTO \end{rotate} & {\bf $\frac{tri. inv}{dtrtri}$}  & \bf 1.90 & \bf 1.40 & \bf 1.26 & \bf 1.15 & \bf 1.10 & \bf 1.07 & \bf 1.08 & \bf 1.05 \\
\hline

\end{tabular}
\caption{Timings of triangular matrix inversion on Itanium2, 1.3GHz}\label{tab:trinv-ia64}
\end{center}
\end{table}

Tables \ref{tab:trinv-p4} and \ref{tab:trinv-ia64} illustrate the performances of our exact triangular matrix inversion regarding performances of LAPACK routine ``dtrtri''.
Results show that our exact computations tend to catch up with the numerical ones and even outperform them
on Itanium2 with ATLAS for large matrices (dimension greater than 8000).

One can notice that the implementation of triangular matrix inversion provided by GOTO is quite efficient 
compare to ATLAS, and thus lead our exact computation to be more efficient but not better than numerical ones.
Here again, this demonstrates that exact triangular matrix inversion over finite field is not much more
costly than its numerical counterpart.


\begin{table}[htb]\begin{center}
\begin{tabular}{|cc|c||r|r|r|r|r|r|r|r|r|}
\cline{3-10}
\multicolumn{2}{c|}{} & $n$        
& {\em 1000}  
& {\em 3000}  & {\em 5000} & {\em 7000}  & {\em 8000}  & {\em 9000} & {\em 10000}\\
\cline{3-10}
\multicolumn{10}{c}{}\\[-0.2cm]
\hline 
& & inverse 
& $0.75$s    
& $13.57$s    &
$54.52$s &  $141.19$s   & $206.26$s  & $285.19$s & $385.35$s \\ 
\cline{3-10}
& & dgetrf+dgetri  
& $0.69$s    
& $16.94$s
&   $80.83$s &  $222.07$s   & $368.66$s    & $531.29$s & $761.28$s \\ 
\cline{3-10} \\[-.3cm]
\cline{3-10}
& \begin{rotate}{90}\scriptsize ATLAS \end{rotate} & {\bf
  $\frac{inverse}{dgetrf+dgetri}$}  
& \bf 1.09  
& \bf 0.80  &  \bf 0.67  &  \bf 0.64  & \bf 0.56  & \bf 0.54 & \bf 0.51  \\   
\hline
\multicolumn{10}{c}{}\\[-0.2cm]
\hline
& & inverse 
& $0.63$s    
& $11.82$s    &
$48.56$s &  $125.30$s   & $179.17$s  &$256.12$s & $343.91$s \\ 
\cline{3-10}
& & dgetrf+dgetri 
& $0.55$s    
& $13.02$s   &
$58.36$s &  $159.21$s   & $232.30$s  & $328.55$s& $450.46$s \\ 
\cline{3-10} \\[-.3cm]
\cline{3-10}
& \begin{rotate}{90}\scriptsize GOTO \end{rotate} & {\bf
  $\frac{inverse}{dgetrf+dgetri}$}  
& \bf 1.15  
& \bf 0.91   &  \bf 0.83  &  \bf 0.79  & \bf 0.77  &  \bf 0.78 &  \bf 0.76  \\   
\hline

\end{tabular}
\caption{Timings of matrix inversion on a Xeon, 3.6GHz}\label{tab:inv-p4}


\begin{tabular}{|cc|c||r|r|r|r|r|r|r|}
\multicolumn{1}{c}{}\\
\cline{3-10}
\multicolumn{2}{c|}{} & $n$          & {\em 1000}    & {\em 3000}  & {\em 5000} & {\em 7000}  & {\em 8000}  & {\em 9000} &  {\em 10000}\\
\cline{3-10}
\multicolumn{10}{c}{}\\[-0.2cm]
\hline
& & inverse& $1.01$s & $17.27$s & $69.24$s & $173.21$s & $256.67$s & $353.02$s & $483.08$s  \\
\cline{3-10}
& & dgetrf+dgetri & $0.60$s & $14.29$s & $66.08$s & $184.74$s & $276.09$s & $393.62$s & $541.37$s  \\
\cline{3-10} \\[-.3cm]
\cline{3-10}
& \begin{rotate}{90}\scriptsize ATLAS \end{rotate} & {\bf $\frac{inverse}{dgetrf+dgetri}$}  & \bf 1.67 & \bf 1.21 & \bf 1.05 & \bf 0.94 & \bf 0.93 & \bf 0.90 & \bf 0.89 \\
\hline
\multicolumn{10}{c}{}\\[-0.2cm]
\hline
& & inverse & $0.85$s & $14.92$s & $61.00$s & $153.78$s & $226.68$s & $313.84$s & $422.78$s  \\
\cline{3-10}
& & dgetrf+dgetri & $0.47$s & $11.45$s & $51.33$s & $139.00$s & $207.36$s & $293.02$s & $402.72$s  \\
\cline{3-10} \\[-.3cm]
\cline{3-10}
& \begin{rotate}{90}\scriptsize GOTO \end{rotate} & {\bf $\frac{inverse}{dgetrf+dgetri}$}  & \bf 1.80 & \bf 1.30 & \bf 1.19 & \bf 1.11 & \bf 1.09 & \bf 1.07 & \bf 1.05 \\
\hline
\end{tabular}
\caption{Timings of matrix inversion on Itanium2, 1.3GHz}\label{tab:inv-ia64}
\end{center}
\end{table}

Now, Tables \ref{tab:inv-p4} and \ref{tab:inv-ia64} provide the same comparisons for dense matrix inversion.
For numerical computation references we use the routine ``dgetri'' in combination with the factorization routine ``dgetrf'' to yield matrix inverse.
On both architecture with ATLAS BLAS kernel, exact computations become the most efficient when matrix dimension is getting larger.
Numerical computation is only better than exact on the Itanium 2 architecture with GOTO BLAS kernel.
In this particular application, the benefit of fast matrix multiplication is important since it allows to outperform numerical performances.\\

\begin{figure}[hbtp]
\begin{minipage}[t]{0.49\textwidth}
\begin{center}
\includegraphics[width=.7\textwidth,height=\textwidth,angle=-90]{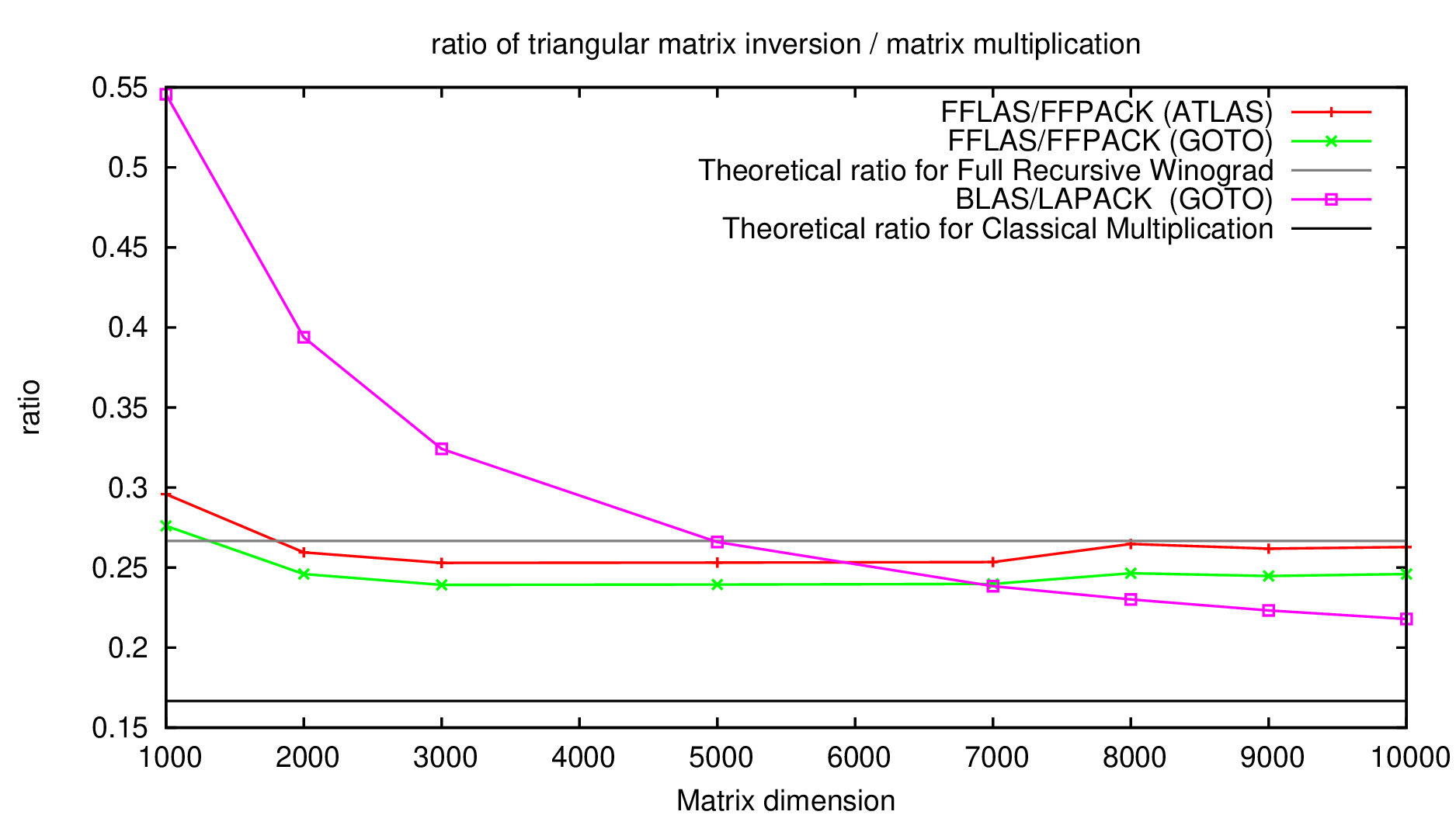}
\end{center}
\caption{Comparing  triangular matrix inversion with matrix
  multiplication on a Xeon,
  3.6GHz}\label{fig:trinv-mul}
\end{minipage}
\hfill
\begin{minipage}[t]{0.49\textwidth}
\begin{center}
  \includegraphics[width=.7\textwidth,height=\textwidth,angle=-90]{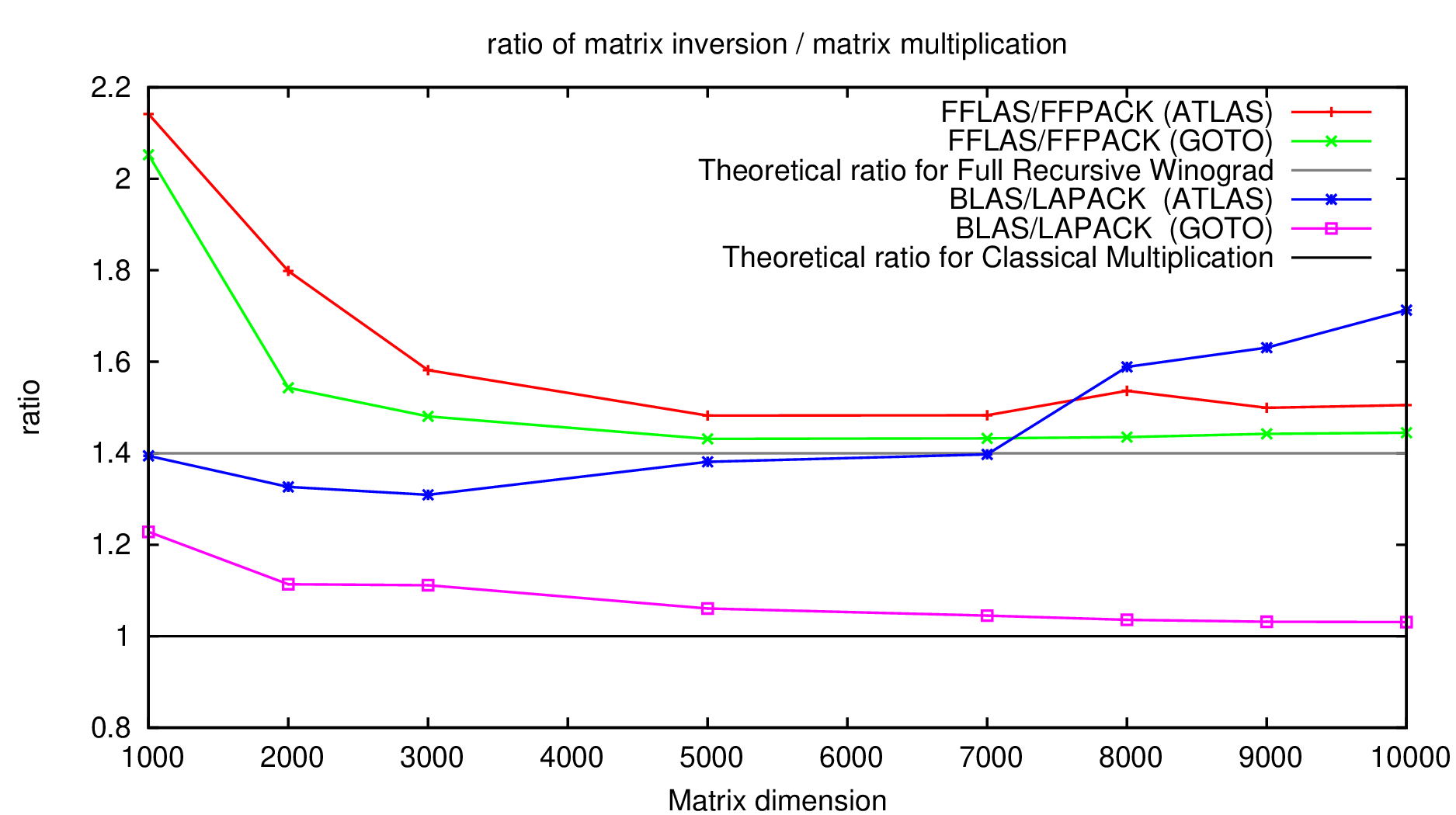}
\end{center}
\caption{Comparing  matrix inversion with matrix multiplication on a Xeon,
  3.6GHz}\label{fig:inv-mul}
\end{minipage}
\end{figure}

As shown in previous section, matrix inversion algorithms reduce to matrix multiplication.
Figures \ref{fig:trinv-mul} and \ref{fig:inv-mul} show the correlation between matrix inversion performances 
and matrix multiplication performances; triangular and dense case are studied.

According to section \ref{sssec:trinverse}, the ratio of triangular matrix inversion and matrix multiplication is $4/(2^\omega-4)(2^\omega-2)$;
which gives a theoretical ratio of $1/6$ when classic matrix multiplication is used.
However this ratio increase to $\approx 0.267$ when Winograd fast matrix multiplication is used (i.e. $\omega = \log_2 7$).
Since our matrix multiplication routine is using fast matrix multiplication, the asymptotic behavior of this ratio should tend to the latter.
However we observe in practice that our performances are beyond this ratio.
This is due to the hybrid matrix multiplication which uses both Winograd and classic algorithms.
So the practical ratio obtained here is really close to the theoretical one since it should asymptotically lie between $0.2674$ and $0.166$.

From section \ref{sssec:inverse} one can express the ratio between dense matrix inversion and matrix multiplication as respectively $1$ 
with classic algorithm and $1.4$ with Winograd algorithm.
In practice we observe that dense matrix inversion ratio is just above the asymptotic behavior of Winograd based inversion.
This certainly could be explained by the number of different algorithms involved in this application.
In particular it involves three different reductions to matrix multiplications; which may be of a little influence on the final performances.
Moreover, we do not take into account memory effect which can play a crucial role in performances as already demonstrated by ATLAS software with optimized BLAS \cite{Whaley:2001:AEO}. In our test we used a naive approach which leads us to use $2n^2$ elements in memory. Decreasing this memory  will certainly allow us to get better performances.
In particular, it is not known yet how to perform matrix inversion in place using a reduction to matrix multiplication.


%% file: concl.tex
\section{Conclusions}
We have achieved the goal of approaching the efficiency of the
numerical linear algebra library but for word-size prime fields.
We showed that exact computation can benefit from Winograd fast matrix multiplication algorithm  and then even leads to 
 outperform the efficiency of the well known BLAS and LAPACK libraries.

This performance is achieved through efficient reduction to matrix multiplication where we took care of minimizing the ratio and also 
by reusing the numerical computation as much as possible.
We also showed that from our routines one can easily implement efficient algorithms for many linear algebra problems (e.g. null-space, generalized inverse, etc.).
Note that approximate timings for these algorithms can be derived from the timings provided with our main routines.


One can try to design block algorithms where the blocks fit in the cache of a specific machine to reach very good efficiency.
By reusing BLAS library this has been proven to be almost useless for matrix multiplication in \cite{jgd:2002:fflas} and we think we proved here that this is 
not mandatory also for any dense linear algebra routine. 
Therefore, using recursive block algorithms, efficient numerical BLAS and fast matrix multiplication algorithms 
one can approach the numerical performance or even surpass them over
some finite fields.
Moreover, long range efficiency and portability are warranted as
opposed to every day tuning. Except for small matrices where the
conversions increase slightly the running time, and except for the LQUP
transform, we have shown that all our exact routines can be faster
than their numerical counterparts. 

Besides, the exact equivalent of stability constraints for numerical computations is coefficient growth. 
Therefore, whenever possible, we computed and improved theoretical
bounds on this growth (e.g. bounds \ref{cor:trsmcentre} and
\ref{cor:centre}). 
Those optimal bounds enable further uses of the BLAS routines.

Further developments include:\\
%
$\bullet~$ The main case where our wrapping of BLAS is insufficient is
for very small matrices where benefits of BLAS are limited and fast
algorithms are not useful. Here, a design using the finite field
directly might improve the speed.\\
$\bullet~$ More generally, a Self-adapting Software \cite{Dongarra:2003:SANS} would allow to provide hybrid implementations with best empirical thresholds.\\
$\bullet~$ The technique of wrapping BLAS becomes useless when finite fields are larger than the corresponding bound of feasibility 
(e.g. $p> 2^{26}$ for matrix multiplication). At a non negligible price the Chinese remainder algorithm could be used to authorize the use of BLAS.
Optimizing this scheme would then be an interesting way to provide
similar results for larger finite fields.\\
$\bullet~$ Finally, extending the out of core versions by more
recursive data format and the building of a parallel library is promising.
Also, in the case of parallelism, our all-recursive approach
enables a very efficient ``sequential-first'' parallelization as shown
e.g. in \cite{jgd:2006:atrsm} for triangular system solving. 

%% file: winobound.tex
\section{Proof of theorem \ref{th:bound}}\label{app:winobound}
Consider the natural block decomposition
$$
\begin{bmatrix}
  C_{11} & C_{12}\\
  C_{21} & C_{22}\\
\end{bmatrix}
=
\begin{bmatrix}
  A_{11} & A_{12}\\
  A_{21} & A_{22}\\
\end{bmatrix}
\begin{bmatrix}
  B_{11} & B_{12}\\
  B_{21} & B_{22}\\
\end{bmatrix},
$$
where $A_{11}$ and $B_{11}$ have respectively dimension $m/2 \times k/2$ and $k/2
\times n/2$.

To bound the intermediate values in the computation of $l$ recursive
levels of Winograd's algorithm, we will show that the worst case
occurs in the computation of one of the intermediaite products.
We will first consider the case $K=2^lq$ and then generalize the
result for every $K$. To end the proof we will provide an instance of
a computation for which the bound is attained.

\subsection{Some properties on the series of the type $2u-v$}

Consider the series defined recursively by:
\[ \left\{ 
\begin{array}{l} 
u_{l+1} = 2u_l-v_l \\ v_{l+1} = 2v_l-u_l\\
u_{0} \leq 0\\ v_{0} \geq 0 \\
\end{array}
\right.
\]
Since
\[\left\{
\begin{array}{l}
u_{l+1}+v_{l+1}=u_{l}+v_{l}=\dots=u_{0}+v_{0}\\
v_{l+1}-u_{l+1}=3(v_{l}-u_{l})=\dots=3^{l+1}(v_{0}-u_{0})\\
\end{array}
\right.
\]
It comes
\[ \left\{ 
\begin{array}{l} 
u_{l} = u_0\frac{(1+3^l)}{2}+v_0\frac{(1-3^l)}{2}\\
v_{l} = v_0\frac{(1+3^l)}{2}+u_0\frac{(1-3^l)}{2}
\end{array}
\right.
\]
Thus, the following properties hold:
\begin{align}
& u_l \leq 0 ~\text{and}~ v_l \geq 0\\
& u_{l}  ~\text{is decreasing and}~ v_{l} ~\text{ is increasing} \label{eq:inc}\\
& v_l > -u_l  ~\text{if}~ v_0 > -u_0\label{eq:sup}
\end{align}

Now define $v^A$ and $v^B$, two series of the type $v$ by setting
$u^A_0=m_A$, $v^A_0=M_A$, $u^B_0=m_B$ and $v^B_0=M_B$.

Let us also define 
$ t_j = \frac{1+3^j}{2}$ and  $s_j = \frac{1-3^j}{2}$. Thus $t_j + s_j = 1$
and $t_j - s_j = 3^j$.

The following property holds:
\begin{equation}\label{eq:vap}
 (2M_A-m_A)t_j+(2m_A-M_A)s_j = M_At_{j+1}+m_As_{j+1} = v^A_{j+1}
\end{equation}

\subsection{Notations}\label{sec:notations}

Let 
$$
b_l =
\left(\frac{1+3^l}{2}M_A+\frac{1-3^l}{2}m_A\right)\left(\frac{1+3^l}{2}M_B+\frac{1-3^l}{2}m_B\right) 
  \left\lfloor{ \frac{K}{2^l}}\right\rfloor.
$$
The serie $(b_l)_{l>0}$ is increasing since (\ref{eq:inc}).

Winograd's implementation, see
e.g. \cite{Huss-Lederman:1996:mai,DumasPernet:2007:WinoSchedule},
uses the following intermediate computations
\begin{eqnarray*}
P_1 &=& A_{11} \times B_{11} \\
P_2 &=& A_{12} \times B_{21} + \beta C_{11}\\
P_3 &=& (A_{12}+A_{11}-A_{21}-A_{22}) \times B_{22}\\
P_4 &=& A_{22} \times (B_{22}+B_{11}-B_{21}-B_{12}) + \beta (C_{22}-C_{12}-C_{21})\\
P_5 &=& (A_{21}+A_{22}) \times (B_{12} - B_{11}) + \beta C_{12}\\
P_6 &=& (A_{21}+A_{22}-A_{11}) \times (B_{22}+B_{11}-B_{12}) \\
P_7 &=& (A_{11}-A_{21}) \times (B_{22} - B_{12}) +\beta
(C_{22}-C_{12})\\
C_{11}=U_1&=&P_2+P_1\\
U_2 &=&(A_{21}+A_{22}-A_{11})\times (B_{22}-B_{12})+(A_{21}+A_{22})\times B_{11} \\
U_3 &=& A_{22} \times (B_{22} - B_{12})+(A_{21}+A_{22}) \times B_{11} +\beta (C_{22}-C_{12}) \\
U_4 &=& (A_{21}+A_{22}) \times B_{22} + A_{11} \times (B_{12} -
B_{22})  + \beta C_{12}\\
C_{12}=U_5&=&U_4+P_3\\
C_{21}=U_6&=&U_3-P_4\\
C_{22}=U_7&=&U_3+P_5\\
\end{eqnarray*}

Remark that the result of the computation is independent of the
algorithm and is always bounded by
$K \text{max}(|m_A|,|M_A|)\text{max}(|m_B|,|M_B|) + \beta 
\text{max}(|m_C|,|M_C|) \leq (K+1) M_AM_B$.
Now this value is always smaller than $b_1$ for $k\geq 1$ and also
smaller than $b_l \ \forall l\geq 1$.
Therefore, the coefficients of the blocks $U_1$, $U_5$, $U_6$ and  $U_7$
always satisfy the bound.
Now if the remaining 9 intermediate computations are bounded by
$b_l$, we will be done.

We will prove that the largest intermediate value always occurs in the
computation of $P_6$.
Consider $l$ recursive levels indexed by $j$: $j=l$ is the first
splitting of the matrices into four blocks and $j=0$ corresponds to 
the last level where the product is done by a classic matrix
multiplication algorithm.
The recursive algorithm can be seen as a back and forth process: the
splitting is done from $j=l$ to $j=0$ and then the multiplications are
done from $j=0$ to $j=l$.

We also define the following notations:
\begin{itemize}
\item $M^{j,k}_{m_A,M_A,m_B,M_B,m_C,M_C}(X)$ is an upper bound on the
  intermediate computations of $X=A \times B +\beta C$ with
  $j$ recursive levels and $m_A\leq a_{i,j}\leq M_A$,
  $m_B\leq b_{i,j} \leq M_B$ and $m_C\leq c_{i,j}\leq M_C$.
$k$ is the common dimension of  $A$ and $B$
\item $M^{j,k}_{m_A,M_A,m_B,M_B,m_C,M_C} = \max_{X}
 M^{j,k}_{m_A,M_A,m_B,M_B,m_C,M_C}(X)$.
\item $M(X)\frac{k}{2^{j+1}}$ for $M^{j+1,k}_{m_A,M_A,m_B,M_B,m_C,M_C}(X)$.
\end{itemize}

The following formulas correspond to the seven recursive calls:
\begin{equation}\label{sys:max}
\begin{array}{c}
M^{j+1,k}_{m_A,M_A,m_B,M_B,m_C,M_C} = \\
\max {\left(
        \begin{array}{l}
                M(P_1) = M^{j,\frac{k}{2}}_{m_A,M_A,m_B,M_B,0,0} \\
                M(P_2) = M^{j,\frac{k}{2}}_{m_A,M_A,m_B,M_B,m_C,M_C} \\
                M(P_3) = M^{j,\frac{k}{2}}_{2m_A-2M_A,2M_A-2m_A,m_B,M_B,0,0} \\
                M(P_4) = M^{j,\frac{k}{2}}_{m_A,M_A,2m_B-2M_B,2M_B-2m_B,m_C-2M_C,M_C-2m_C} \\
                M(P_5) = M^{j,\frac{k}{2}}_{2m_A,2M_A,m_B-M_B,M_B-m_B,m_C,M_C} \\
                M(P_6) = M^{j,\frac{k}{2}}_{2m_A-M_A,2M_A-m_A,2m_B-M_B,2M_B-m_B,0,0} \\
                M(P_7) = M^{j,\frac{k}{2}}_{m_A-M_A,M_A-m_A,m_B-M_B,M_B-m_B,m_C-M_C,M_C-m_C} \\
        \end{array}\right)
} 
\end{array}
\end{equation}
Moreover, the classic algorithm is 
used for $j=0$:
\begin{equation}\label{eq:classic}
M^{0,k}_{m_A,M_A,m_B,M_B,m_C,M_C} = \text{max}
\left(
\begin{array}{l}
M_A M_Bk + \beta M_C\\
-m_A M_Bk - \beta m_C\\
-M_A m_Bk - \beta m_C\\
\end{array}
\right)
\end{equation}

\subsection{Some invariants} \label{sec:invariants}

\begin{lem}
The following invariants hold in every recursive call:
\begin{enumerate}
\item  $0 \leq -m_A  \leq M_A, \ 0 \leq -m_B \leq M_B,\ 0\leq -m_C \leq M_C$
\item  $m_C \geq m_B$ and $M_C\leq M_B$
\item  $M_C-m_C \leq M_B-m_B$
\end{enumerate}
\end{lem}

\begin{proof}
From equation (\ref{sys:max}), one gets invariants ($1$) and ($2$). 
Then invariant ($3$) is a consequence of ($1$) and ($2$).
\end{proof}

\subsection{Induction for $K= 2^lq$}\label{sec:recurrence}

Let $IH_j$ be the following induction hypothesis:

\textit{
If the invariants of section \ref{sec:invariants} are satisfied then
$$ M^{j,k}_{m_A,M_A,m_B,M_B, m_C, M_C} = [v^A_j ][v^B_j]\frac{k}{2^j}.
$$
}
Suppose that the previous invariants are satisfied and that $IH_j$ 
is true. We will prove that the maximum of (\ref{sys:max}) is reached during the 
computation of $P_6$ to show that $IH_{j+1}$ is satisfied.

The conditions on $m_A$, $M_A$, $m_B$ and $M_B$ are satisfied for every recursive call.
We can therefore apply $IH_j$ to  every product
$X \in \{P_1,P_2,P_3,P_4,P_5,P_6 \}$ in order to compare
$M(X)$ with $M(P_6)$. 

\begin{itemize}
\item For $P_1 = A_{11} \times B_{11}$:
\begin{eqnarray*}
  M(P_6)-M(P_{1}) & = & \left[ (2M_A-m_A)t_j+(2m_A-M_A)s_j \right]  \times\\
& &  \left[ (2M_B-m_B)t_j+(2m_B-M_B)s_j \right] - v^{A_{11}}_jv^{B_{11}}_j\\
                  & = & v^A_{j+1}v^B_{j+1}-v^{A_{11}}_jv^{B_{11}}_j\\
                  & \geq & v^A_{j+1}v^B_{j+1}-v^A_jv^B_j\\
\end{eqnarray*}
And since $v^A$ and $v^B$ are increasing and positive, we have 
$M(P_6)\geq M(P_{1})$.

\item For $P_2 = A_{12} \times B_{21} + \beta C_{11}$: with the same argument
$M(P_6)\geq M(P_2)$.

\item For $P_3 = (A_{12}+A_{11}-A_{21}-A_{22}) \times B_{22}$:
\begin{eqnarray*}
  M(P_6)-M(P_3) & = & v^A_{j+1}v^B_{j+1}-v^{A_{11}+A_{12}-A_{21}-A_{22}}_jv^{B_{22}}_j  \\
                & = & v^A_{j+1}v^B_{j+1}-[(2M_A-2m_A)t_j+(2m_A-2M_A)s_j]v^B_j \\
                & = & v^A_{j+1}v^B_{j+1}-(v^A_{j+1}-m_A t_j-M_A s_j  )v^B_j (\ref{eq:vap}) \\
                & = & v^A_{j+1}[v^B_{j+1}-v^B_j] - u^A_jv^B_j  \\
                &\geq& v^A_{j+1}[v^B_{j+1}-v^B_j] - v^A_{j+1}v^B_j (\ref{eq:sup})\\
                &\geq& v^A_{j+1}[v^B_{j+1}-2v^B_j]\\
                &\geq& v^A_{j+1}3^j[M_B-m_B]\geq 0
\end{eqnarray*}

\item For $P_4= A_{22} \times (B_{22}+B_{11}-B_{21}-B_{12}) 
           + \beta (C_{22}-C_{12}-C_{21})$: with the same argument,
\begin{equation*}
M(P_6)-M(P_4)  = v^A_{j+1}v^B_{j+1}-v^{A_{22}}_jv^{B_{22}+B_{11}-B_{12}-B_{21}}_j \geq 0
\end{equation*}

\item For $P_5 = (A_{21}+A_{22}) \times (B_{12} - B_{11}) + \beta C_{12}$:
\begin{eqnarray*}
 M(P_6)-M(P_5) & = & v^A_{j+1}v^B_{j+1}-v^{A_{21}+A_{22}}_jv^{B_{12}-B_{11}}_j\\
               & = & v^A_{j+1}v^B_{j+1}-2v^A_j\left[v_j^B - u_j^B\right]\\
               & = & \left[2v^A_j-u^A_j\right]v^B_{j+1}-v^A_j\left[v^B_{j+1} - u_j^B \right]\\
               & = & v^A_j v^B_{j+1} -u^A_j v^B_{j+1} + v^A_j u_j^B\\
               & = & v^A_j\left[ v^B_{j+1} + u_j^B \right] -u^A_j v^B_{j+1} \\
               & = & v^A_j\left[ 2 v_j^B \right] -u^A_j v^B_{j+1} 
\end{eqnarray*}
and since $u^A_j \leq 0 \leq v^A_j,  v_j^B, v^B_{j+1}$ it comes
 $M(P_6)-M(P_5) \geq 0$.

\item For  $P_7 = (A_{11}-A_{21}) \times (B_{22} - B_{12}) + \beta (C_{22}-C_{12})$:
using $P_5$,
\begin{eqnarray*}
  M(P_5)-M(P_7) & = & v^{A_{21}+A_{22}}_jv^{B_{12}-B_{11}}_j-v^{A_{11}-A_{21}}_jv^{B_{22}-B_{12}}_j\\
                & = & [ 2M_At_j+2m_As_j-(M_A-m_A)t_j-(m_A-M_A)s_j  ] \times\\
                &   &  \left[(M_B-m_B)t_j+(m_B-M_B)s_j\right]\\
                & = & \left[(M_A+m_A)(t_j+s_j)\right]\left[(M_B-m_B)(t_j-s_j)\right]\\
                &\geq& 0
\end{eqnarray*}

The coefficients of the blocks $U_1, U_5, U_6$ and $U_7$ are bounded by 
$kM_AM_B + \beta M_C$ and are therefore smaller than the ones in $P_6$.

Lastly, we must control the size of the coefficients in  $U_2=P_1+P_6$, 
$U_3=U_2+P_7$ and $U_4=U_2+P_7$. 

\item For $U_2 = (A_{21}+A_{22}-A_{11}) \times (B_{22}-B_{12})
           +(A_{21}+A_{22}) \times B_{11}$:
\begin{equation}\label{eq:u2}
\forall x \in U_2, |x|\leq \max{\left(
    \begin{array}{l} 
      (2M_A-m_A)(M_B-m_B)+2M_AM_B\\
      (-2m_A+M_A)(M_B-m_B)-2m_AM_B\\
      (-2m_A+M_A)(M_B-m_B)-2M_Am_B
    \end{array}\right)}k/2^j
\end{equation}
Now 
$ 2M_A-m_A-(-2m_A+M_A)=M_A+m_A \geq 0$ and $ 0\leq -m_A \leq M_A$, so 
the \ref{eq:u2} simplifies into 
$\forall x \in U_2, |x|\leq (2M_A-m_A)(M_B-m_B)+2M_AM_B$.

\begin{eqnarray*}
M(P_6) - M(U_2) & \geq & (2M_A-m_A)(2M_B-m_B) - (2M_A-m_A)(M_B-m_B)\\
                & & -2M_AM_B\\
                & = & (2M_A-m_A)(M_B) - 2M_AM_B\\
                & = & -m_AM_B \geq 0
\end{eqnarray*}

\item For $U_3 = A_{22} \times (B_{22} - B_{12})+(A_{21}+A_{22}) \times B_{11} 
           + \beta(C_{22}-C_{12})$:
with the same argument
\begin{equation*}
\forall x \in U_3, |x|\leq \max{\left(
    \begin{array}{l} 
      (M_A(M_B-m_B)+2M_AM_B)k/2^j+|\beta|(M_C-m_C)\\
      (M_A(M_B-m_B)-2m_AM_B)k/2^j+|\beta|(M_C-m_C)\\
      (M_A(M_B-m_B)-2M_Am_B)k/2^j+|\beta|(M_C-m_C)\\
    \end{array}\right)}k/2^j
\]
The max is always equal to its first argument, and since
$ k/2^j \geq 1$, $\beta \leq M_A-m_A$ and $ M_C-m_C \leq M_B-m_B$, we have:
\begin{eqnarray*}
  |x| &\leq & (M_A(M_B-m_B)+2M_AM_B)k/2^j+\beta(M_C-m_C)  \\
      &\leq &(2M_A-m_A)(M_B-m_B)+2M_AM_B)k/2^j  \\
      &\leq &M(U2)\leq M(P6) 
\end{eqnarray*}


\item For $U_4=(A_{21}+A_{22}) \times B_{22} + A_{11} \times (B_{12}
  - B_{22})  + \beta C_{12}$: 
with the same argument as for $U_3$,

\begin{equation*}
\forall x \in U_4, |x|\leq (M_A(M_B-m_B)+2M_AM_B)k/2^j+|\beta| M_C
\end{equation*}

Since $M_C \leq M_B -m_B$, $-m_A\leq M_A$ and $-m_B\leq M_B$, 
we have $$M(U_4)\leq M(U_3) \leq M(P_6).$$
\end{itemize}

Finally $M^{j+1,k}_{m_A,M_A,m_B,M_B} =
M(P_6)\frac{k}{2^{j+1}} = v^A_{j+1}v^B_{j+1}\frac{k}{2^{j+1}}$, and
$IH_{j+1}$ is satisfied.

\vspace{1em}

For the initialization of the induction ($j=1$), the products of the blocks
are done by the classical algorithm. 
From (\ref{sys:max}) and (\ref{eq:classic}), one gets:
{\small
\begin{eqnarray*}
M^{1,k}_{m_A,M_A,m_B,M_B,m_C,M_C}(P_1) & = &  M_A M_Bk/2\\
M^{1,k}_{m_A,M_A,m_B,M_B,m_C,M_C}(P_2) & = & M_A M_Bk/2 + |\beta| M_C\\
M^{1,k}_{m_A,M_A,m_B,M_B,m_C,M_C}(P_3) & = & 2(M_A-m_A) M_Bk/2\\
M^{1,k}_{m_A,M_A,m_B,M_B,m_C,M_C}(P_4) & = & 2M_A (M_B-m_B)k/2 + |\beta| (2M_C-m_C)\\
M^{1,k}_{m_A,M_A,m_B,M_B,m_C,M_C}(P_5) & = & 2M_A (M_B-m_B)k/2 + |\beta| M_C\\
M^{1,k}_{m_A,M_A,m_B,M_B,m_C,M_C}(P_6) & = & (2M_A-m_A)(2M_B-m_B)k/2\\
M^{1,k}_{m_A,M_A,m_B,M_B,m_C,M_C}(P_7) & = & (M_A-m_A)(M_B-m_B)k/2+|\beta|(M_C-m_C)\\
M^{1,k}_{m_A,M_A,m_B,M_B,m_C,M_C}(U_2) & = & (2M_A-m_A)( M_B-m_B)k/2+2M_AM_Bk/2\\
M^{1,k}_{m_A,M_A,m_B,M_B,m_C,M_C}(U_3) & = & M_A (M_B-m_B)k/2+2M_A M_Bk/2+|\beta|(M_C-m_C)\\
M^{1,k}_{m_A,M_A,m_B,M_B,m_C,M_C}(U_4) & = & 2M_A M_Bk/2+M_A(M_B-m_B)k/2+|\beta| M_C\\
\end{eqnarray*}
}
Again, we will prove that $M^{1,k}_{m_A,M_A,m_B,M_B,m_C,M_C}(P_6)$ reaches the
highest value, using invariants of section \ref{sec:invariants}, and the fact that
$|\beta| \leq M_A, M_B$ and $k\geq 2$.

It is straightforward for $P_1$ and $P_2$. 
\begin{itemize}
\item For $P_3$:
{\small
\begin{eqnarray*}
M^{1,k}_{m_A,\dots}(P_6)-M^{1,k}_{m_A,\dots}(P_3)
 &=& ((2M_A-m_A)(2M_B-m_B)-2(M_A-m_A) M_B)k/2\\
 &=& (2M_AM_Bk-(2M_A-m_A)m_B)k/2 \geq 0
\end{eqnarray*}
}
\item For $P_4$: Since $-|\beta|(2M_C-m_C)\geq-M_A(2M_B-m_B)$, we have
{\small
\begin{eqnarray*}
M^{1,k}_{m_A,\dots}(P_6)-M^{1,k}_{m_A,\dots}(P_4) &=& ((2M_A-m_A)(2M_B-m_B)-2M_A(M_B-m_B))k/2 \\
                                                  & &-|\beta|(M_C-2m_C) \\
                                                  &\geq& (M_A-m_A)(2M_B-m_B)-2M_A(M_B-m_B)\\
                                                  &=& m_A(m_B-2M_B) \geq 0
\end{eqnarray*}
}
\item For $P_5$: $M^{1,k}_{m_A,M_A,m_B,M_B,m_C,M_C}(P_5)\leq M^{1,k}_{m_A,M_A,m_B,M_B,m_C,M_C}(P_4)$

\item For $P_7$: 
{\small
\begin{eqnarray*}
M^{1,k}_{m_A,\dots}(P_6)- M^{1,k}_{m_A,\dots}(P_7) 
&=&  ((2M_A-m_A)(2M_B-m_B) \\
& &-(M_A-m_A)(M_B-m_B)k/2 -|\beta|(M_C-m_C)\\
&\geq& M_A(2M_B-m_B)+(M_A-m_A)M_B-M_A(M_B-m_B)\\
&\geq& (2M_A-m_A)M_B \geq 0
\end{eqnarray*}
}
\item For $U_2$, $U_3$, $U_4$: using the same argument as for the case of arbitrary $j$.
\end{itemize}
$IH_1$ is then satisfied.

\subsection{Case of an arbitrary $k$}

Let $l$ be such that $ 2^l d \leq k <  2^l(d+1)$ ( $d=\left\lfloor \frac{k}{2^l} \right\rfloor$).
A \textit{dynamic peeling} technique \cite{Huss-Lederman:1996:ISA}
is used to deal with odd dimensions:
at each recursive level, the largest blocks with even dimensions at the top left
hand corner of the input matrices are multiplied using Winograd's algorithm.
Then an optional rank $1$ update is applied, with the odd dimensions.

These updates are using matrix-vector products, dot products and tensor products.
Every intermediate result during these computations are therefore bounded in absolute
value by $ k M_AM_B + |\beta|M_C \leq (k+1)M_AM_B$

We show now that this bound is always under the one of Winograd's algorithm. 
$$
\forall l\geq 1 \ 2^l(d+1)M_AM_B \leq 
v_l^Av_l^B\left\lfloor\frac{k}{2^l}\right\rfloor 
$$
(since $ (k+1)M_AM_B \leq 2^l(d+1)M_AM_B $).

\begin{itemize}
\item For $l=1$, the inequation is satisfied: $2M_AM_B(d+1) \leq
  (2M_A-m_A)(2M_B-m_B)d$ (since $d\geq 1$)
\item Let us suppose that it is satisfied for $l\geq 1$ and prove it for $l+1$:
\begin{eqnarray*}
v_{l+1}^Av_{l+1}^B\left\lfloor\frac{k}{2^{l+1}}\right\rfloor &=&
[(2M_A-m_A)t_l+(2m_A-M_A)s_l]\\
&\times&[(2M_B-m_B)t_l+(2m_B-M_B)s_l]d\\
&\geq & 2[M_At_l+m_As_l][M_Bt_l+m_Bs_l]2d\\
&\geq & v_{l}^Av_{l}^B\left\lfloor\frac{k}{2^{l}}\right\rfloor\\
&\geq & 2(2^lM_AM_B(2d+1))\\
&\geq & 2^{l+1}M_AM_B(d+1))
\end{eqnarray*}
\end{itemize}
By induction, the bound of section \ref{sec:recurrence} is valid for any $k$.

\subsection{Optimality of the bound}

We simply build a sequence of square matrices $A_l$ and $B_l$ of order $2^l$ for which $l$ recursive 
calls to Winograd's algorithm will involve intermediate results equals to the bound.

Let $(A_l)_{l \in \mathbb{N}^*}$ and $(B_l)_{l \in \mathbb{N}^*}$ be recursively defined as follows:
\[ 
\left\{
\begin{array}{rr}
A_1 = \left[ \begin{array}{cc} m_A & 0 \\ M_A & M_A \end{array} \right], &
B_1 = \left[ \begin{array}{cc} M_B & m_B \\ 0 & M_B \end{array} \right]  \\
&\\
A_{l+1}=\left[\begin{array}{cc}\overline{A_l}&0\\A_l&A_l\end{array}\right], &
B_{l+1}=\left[\begin{array}{cc}B_l&\overline{B_l}\\0&B_l\end{array}\right]
\end{array}
\right.
\]
where $ \overline{A_{i,j}} = M_A+m_A-A_{i,j} $ and $ \overline{B_{i,j}} = M_B+m_B-B_{i,j} $.

Since at each recursive level, the computation of $P6=(A_{21}+A_{22}-A_{11})\times(B_{22}+B_{11}-B_{12})$
involves the largest possible intermediate values, let us define:
\[S(A_l)= (A_l)_{2,1} +(A_l)_{2,2}-(A_l)_{1,1} = 2  A_{l-1} - \overline{A_{l-1}} = 3  A_{l-1} - J _{l-1}\]
where $J_k$ is the square matrix of order $2^k$ 
whose coefficients are all equals to $M_A+m_A$.

Moreover $S(J_k) = J_{k-1}$.
Thus, applying $P_6$  $l$ times recursively, since $S$ is linear:
\[ S(S(\ldots (S(A_l)))) = S^l(A_l) = 3^{l-1} S(A_1) -
\left( \sum_{k=0}^{l-2} 3^k \right) J_1 
\]
Then $S(A_1) = 2M_A-m_A$ and $J_1 = M_A+m_A$ imply:
$$
S^l(A_l) = 3^{l-1} (2M_A-m_A) - \frac{3^{l-1}-1}{3-1}(M_A+m_A) = \frac{1+3^l}{2}M_A+ \frac{1-3^l}{2}m_A .
$$
The same holds for $B_l$:
$$
S^l(B_l) =  \frac{1+3^l}{2}M_B+ \frac{1-3^l}{2}m_B
$$

The order of $A_l$ and $B_l$ is $k=2^l$, so $\left\lfloor\frac{k}{2^l}\right\rfloor=1$.
Therefore, the computation of $ A_l \times B_l$ with $l$ recursive levels of Winograd's algorithm involves
intermediate values equals to $v_l^{A_l}v_l^{B_l}\left\lfloor\frac{k}{2^l}\right\rfloor$. This proves the optimality 
of the bound.

Note that this bound is unchanged for computations of the type $A\times B +\beta C$.